\documentclass[lettersize,journal]{IEEEtran}
\usepackage{graphicx} %graphics
\usepackage{amsmath,bm,mathtools} %Equations
\usepackage{amsmath,amssymb,amsfonts}
\usepackage{bbm}
\usepackage{cite}
\usepackage{colortbl}
\usepackage{subcaption}
\usepackage[shortlabels]{enumitem}
\usepackage{float}
\usepackage{breqn}
\usepackage{lipsum,graphicx}
\usepackage{cite}
\usepackage{algorithm}
\usepackage{algorithmic}
\usepackage{lipsum}
\usepackage{mathtools}
\usepackage{cuted}
\usepackage{stfloats}
\usepackage{siunitx}

\usepackage{scalerel,stackengine}
\usepackage{fontenc}
\usepackage{multicol}
\usepackage{multirow}
\usepackage{xcolor}
\usepackage{diagbox}
\usepackage{soul}
\usepackage{xcolor}
\usepackage{cuted} 
\DeclareMathOperator*{\argmin}{arg\,min}
\usepackage{graphicx}
\usepackage[normalem]{ulem}

\begin{document}
 \title{Channel Estimation for OTFS Systems with Overspread Doppler Shifts}
	\author{Preety Priya,~\IEEEmembership{Member,~IEEE}, Yi Hong,~\IEEEmembership{Senior Member,~IEEE}, and \\Emanuele Viterbo, ~\IEEEmembership{Fellow,~IEEE}
 \thanks{ Preety Priya is with the
Department of Electronics Engineering, IIT (ISM) Dhanbad, Dhanbad 826004, India (e-mail:
preetypriya@iitism.ac.in).}
 \thanks{Yi Hong and Emanuele Viterbo are with the Department of Electrical and Computer Systems Eng, Monash University, Clayton, VIC 3800,
Australia (e-mails: yi.hong@monash.edu, emanuele.viterbo@monash.edu).}
		\thanks{This research work is supported by the Australian Research Council (ARC) through the Discovery project: DP260101376. The work of P. Priya was
supported in part by the Faculty Research Scheme (FRS), IIT (ISM) Dhanbad, under
Grant MISC 0279.}	
	}
\markboth{IEEE Transactions on Wireless Communications,~Vol.~XX, No.~XX, XXX~2025}%
{Shell \MakeLowercase{\textit{et al.}}: A Sample Article Using IEEEtran.cls for IEEE Journals}

{}
\maketitle
\begin{abstract}
%Emerging 6G technologies are envisioned to support ultra-high mobility scenarios. In such environments, Doppler shifts of some propagation paths can be overspread, exceeding half the subcarrier spacing, leading to overspread channel. 

In this paper, we consider an orthogonal time frequency space (OTFS) system in time-varying channels with overspread Doppler shifts, typically found in non-terrestrial
multi-satellite links. The overspread Doppler
shifts with magnitude greater than half of the subcarrier
spacing, result in {\em aliased Doppler shifts} in the delay-Doppler (DD) domain due
to the OTFS modulo operation. This makes channel estimation
very challenging and the traditional channel estimation methods become ineffective.
To address this challenge, we propose a DD training frame and a two-stage channel estimation method. The training frame comprises a cosine pilot signal and a pilot symbol. In the first stage of the channel estimation, the  pilot symbol in the DD domain is utilized to estimate the delays, aliased Doppler shifts, and channel gains of the propagation paths. In the second stage, the received time domain signal is converted into the frequency domain to detect the peaks of all the Doppler shifts  using the cosine pilot signal. Then, we present a threshold-based method to pair the estimated actual
Doppler shifts with their corresponding delays and channel gains. The complexity of the proposed channel estimation is also discussed. Finally, the performance
of the proposed channel estimation is validated in terms of
the normalized mean square error (NMSE) and bit error rate
(BER) in various scenarios.
  \end{abstract}
\IEEEpeerreviewmaketitle
\section{Introduction}
The evolution from the fifth generation (5G) to the sixth generation (6G) wireless technologies is anticipated to bring a substantial advancement in network capabilities, particularly in mobility support, extending coverage from high to ultra-high mobility scenarios such as low Earth orbit (LEO) satellite communications and low-altitude flying platforms \cite{wang2023road6g}. In a high-mobility communication environment, wireless channels can be characterized as time-varying channels with multipath delays and severe Doppler shifts. Here, traditional OFDM waveforms are less effective. Recent advances have shown that orthogonal time frequency space (OTFS) modulation can provide robust performance in such highly dynamic environments, particularly to combat severe Doppler shifts \cite{hadani2017orthogonal,delay_Doppler_book}. The key idea of OTFS is to multiplex information symbols in the delay-Doppler (DD) domain, where the sparse nature of the~
time-varying wireless channels can be captured \cite{delay_Doppler_book}. This operation effectively spreads information symbols throughout the time/frequency resources and enables the
OTFS waveform to achieve maximum effective channel diversity \cite{EffDiversityLett_Ravi}. 
As a result, OTFS converts the time-frequency selective channel (caused by mobility and multiple propagation paths) into a time-independent two-dimensional (2D) channel in the DD domain \cite{tharaj_universal_MRC}. The above merits, and particularly the channel sparsity, can effectively reduce the complexity in channel estimation and equalization, and detection. %

%detection methodologies have been explored for OTFS systems. These include low complexity linear detectors such as LMMSE and zero forcing detectors \cite{Tiwari_Detection,surabhi2019low_comp_lin_equal, pandey2021low,Zou_2021_ICC,naikoti2021signal_choks,our book}, nonlinear detectors like  message passing-based detection \cite{raviteja2018interference} and its variants {\color{red} cite Cholk white paper and Yao ge white paper }, neural network-based detection \cite{DNN_Cholk,CNN_Enku,Calderbank_learning}, and iterative maximal ratio combining based detection \cite{tharaj_rake_MRC_journal,tharaj_universal_MRC}. The authors have also explored oversampled receiver based detection \cite{ge2021receiveroversampling,PP_Oversam_MRC} to enhance the performance compared to the Nyquist sampled receiver.
 
Channel estimation for OTFS systems has been widely investigated in the literature (see \cite{embedded_pilot_raviteja,Chocks_multi_user_MIMO_ch_est_2020,Ravi_Radar,tharaj_OTSM,Thomas_ChaEst,Mishra_superimposed_pilots_ch_est_2022, jesbin_superimposed,learning_choks,Shen_OMP_2019,Liu_uplink_MIMO_ch_est_2020,Shi_TWC_OTFS,Zhao_SBL_2020,Suraj_BSBL_2021,LiXiangjun2025SBLseg,YouGe2025OTFSvariants} and references therein). Different channel estimation approaches have been developed, including the use of embedded pilot(s)  \cite{embedded_pilot_raviteja,Chocks_multi_user_MIMO_ch_est_2020,Ravi_Radar,tharaj_OTSM,Thomas_ChaEst,VihanOTFSISAC2025_embeddedpilots}, superimposed pilots \cite{Mishra_superimposed_pilots_ch_est_2022, jesbin_superimposed}, interleaved pilots \cite{learning_choks,SaifChok2025interleavedCE}, structured training sequences \cite{Shen_OMP_2019,Liu_uplink_MIMO_ch_est_2020,Shi_TWC_OTFS}, expectation-maximization \cite{RIS_OTFS_Yao}, and sparse Bayesian learning-based channel estimation \cite{Zhao_SBL_2020,Suraj_BSBL_2021,LiXiangjun2025SBLseg}.

% %{\color{red} FIX THIS PARAGRAPH MAYBE NEEDS TO BE MOVED
% particularly  in low-latency high mobility
% communications, where the system has a fixed bandwidth and
% a very limited frame size has been investigated, which  in low-latency and ultra-high mobility applications, or overspread Doppler shifts in non-terrestrial communications involving the multi-satellites cooperative transmissions.
% }

Most prior work on OTFS channel estimation typically considered
{\em underspread channels}, where the delay and Doppler shifts do not exceed the block duration and half of the subcarrier spacing, respectively. 
However, a channel can become {\em overspread} in {\em delay} or {\em Doppler}, depending upon the application requirements on data rate and latency. 

For example, our previous work \cite{PriyaLargeDelay2024} has investigated an OTFS system over a channel with {\em overspread delays}, typical of low-latency communications, where the system must operate with very short frames, within a given bandwidth. We found that the overspread delay results in aliased delays due to the OTFS modulo operation in the DD domain. For this case, we proposed a multi-stage channel estimation scheme using a DD domain pilot in conjunction with a time-domain dual-chirp, where the pilot first provides an initial channel estimation to identify the distinct underspread paths, then a dual-chirp correlation is used to obtain the estimated actual delays from the aliased ones, and the other channel parameters of each path.
%Few works \cite{Mehrotra2024ultrahighDoppler, Wang2022LeoCE} have targeted the ultra high mobility scenarios and developed a channel estimation algorithm suitable for it. However, they still assume the Doppler shifts satisfy the underspread condition. 

% But, in low latency ultra high mobility applications with fixed bandwidth and limited frame duration, if we increase the block duration to meet the underspread condition on the delays, the subcarrier spacing has to decrease. When half the subcarrier spacing reduces below the Doppler shifts values of the paths, it results in overspread Dopplers. This consequently leads to aliased Doppler shifts in the DD domain due to the involved modulo operation in the DD domain. The traditional CE fails under such settings as it estimates only the aliased Doppler shifts and not the actual Doppler shifts. Moreover, multiple paths with same delay and aliased Doppler shifts are indistinguishable in DD domain and cannot be estimated with the traditional CE for OTFS systems.

In this paper, we focus on an OTFS system experiencing {\em overspread Doppler shifts}, typically found in non-terrestrial communications involving multi-satellite cooperative transmission \cite{LuoJiaHe_VLEO}. Overspread Doppler shifts occur, due to the simultaneous transmission of the same signals from multiple satellites with largely varying velocities depending on the different satellites directions of motion. 

  Overspread Doppler can also occur due to the Doppler squint in high-mobility wideband systems, where the Doppler shift becomes frequency dependent across subcarriers, leading to an additional Doppler spread proportional to the signal bandwidth \cite{Doppler_squint}. However, in this work, we consider a narrowband OTFS system model and therefore do not account for Doppler squint effects.

The overspread Doppler shifts (with magnitude greater than half of the subcarrier spacing) result in {\em aliased Doppler shifts} in the DD domain due to the OTFS modulo operation. This makes channel estimation very challenging with embedded pilots within data frames.
Assuming stationarity of the DD channel response over a time span of multiple frames, a dedicated pilot frame can be sent periodically to address the challenge. 

The dual chirp time domain correlation-based methods in \cite{PriyaLargeDelay2024} are not applicable for overspread Doppler estimation, since correlation operations fundamentally estimate time shifts rather than frequency shifts.

To address the overspread Doppler channel estimation, we develop a novel training frame scheme and propose a two-stage channel estimation. Specifically, the training frame comprises a cosine pilot signal in addition to a high-power pilot symbol in the DD domain.   The cosine pilots are specifically chosen to facilitate overspread Doppler estimation, because their FFT produces distinct spectral peaks at the actual Doppler shifts. 
In the first stage of channel estimation, the pilot is used to estimate the delays, the aliased Doppler shifts, and the channel gains of the paths. In the second stage, the received time domain signal is converted into the frequency domain to detect the peaks of all the Doppler shifts using the cosine pilot signal. Then, we provide a threshold-based method to pair the estimated actual Doppler shifts with their corresponding delays and channel gains. % $-$ if the gaps between the absolute values of the received signal spectrum samples and the estimated channel gains are. 
The complexity of the proposed channel estimation is also discussed. Finally, the performance of the proposed channel estimation is validated in terms of the normalized mean square error (NMSE) and bit error rate (BER) in various scenarios.

%a mathematical relationship between the channel gains and the amplitude of the received signal spectrum that helps to 

The rest of the paper is organized as follows. Section \ref{Sec:System}
describes the system model. In Section \ref{Sec:PropCE}, a two-stage CE scheme is proposed for channels with overspread Doppler shifts, followed by results and discussions in Section \ref{Sec:Simulation}. Finally, conclusions are drawn in Section~\ref{Sec:Concl}.
 
\indent \textbf{Notations:}  $x, \mathbf{x}, \mathbf{X}$ represent a scalar, a vector, and a matrix, respectively; $(\cdot)^H$, $(\cdot)^T$, $\lvert \cdot \rvert$, $[\cdot]_M$ denote the Hermitian, transpose, modulus, modulo $M$, respectively; $\mathbf X[m,n]$ represents element in $m^{\text{th}}$ row $n^{\text{th}}$ column of matrix $\mathbf X$;  and $\mathbf x[n]$ denotes $n^{\text{th}}$ element of vector $\mathbf x$; $\text{vec}(\mathbf A)$ denotes the column vectorization of matrix $\mathbf A$; $\text{vec}^{-1}(\mathbf a)$ denotes the inverse operation of vectorization; $\mathcal B$ represents a set and $|\mathcal B|$ is the cardinality of the set; $\jmath$ represents unit imaginary number, defined as $\sqrt{-1}$;
%$\boldsymbol 0_M$ is a vector of size $M\times 1$ containing all zeros; 
$\mathbf F_M$ is the $M$-point normalized discrete Fourier transform (DFT) matrix with entries
$\mathbf F_M[m,n]=\frac{1}{\sqrt M}e^{-\jmath2\pi  m n/M}$; $\mathcal O(\cdot)$ represents the Big-$\mathcal O$ notation, and $\mathbb E[\cdot]$ is the expectation operator.
%$\mathcal U(a,b)$ for $a<b \in{\mathbb R}$, denotes the uniform distribution of a random variable (RV) $X$ with a probability density function (pdf) 
% \[
% f_X(x) = \left\{\begin{array}{ll}
%  \frac{1}{b- a} & a\le x \le b \\
% 0 & \text{otherwise} .
% \end{array}\right.
% \]
%The operator ${\bf a}\circ {\bf b}$ represents element-wise multiplication of two same-size vectors ${\bf a}, {\bf b}$.

%%%%%%%%%%%%%%%%%%%%%%%%%%%%%%%%%%%%%%%%%%%
\section{System Model}\label{Sec:System}
We consider an OTFS system operating in a multipath time-varying channel affected by very high Doppler shifts. 

Let us assume that the channel has $L$ paths with a maximum delay $\tau_{\rm max}$ and a maximum Doppler shift $\nu_{\rm max}$ among all paths. The OTFS frame is composed of $N$ time slots (blocks), each with $M$ samples, for a total of $MN$ samples per frame. Assuming the subcarrier spacing $\Delta f$ and the block duration $T=\frac{1}{\Delta f}$, each OTFS frame has the frame duration $T_f=NT$ and occupies a bandwidth $B=M\Delta{f}$.

In the DD domain, we represent the $i$-th path of the time-varying channel, for $i\in[0,L-1]$, by a complex gain $h_i$, a normalized delay $\tau_i=\frac{l_i}{M \Delta f},$ and the normalized Doppler shift $ \nu_i=\frac{k_i}{N T}$, and the {\em normalized maximum delay/Doppler shifts} are $l_{\rm max}=M \Delta f \tau_{\rm max}$  and $k_{\rm max}=NT \nu_{\rm max}$, respectively. 

Note that, throughout the paper, we assume sufficiently high delay and Doppler resolutions, thus we only consider the {\em integer normalized delay and Doppler shifts}.

Unlike the underspread channel $(\tau_{\rm max}\nu_{\rm max}<1)$, this work assumes  an {\em overspread channel} ($\tau_{\rm max}\nu_{\rm max}>1$) with very high Doppler shifts, i.e., 
\[
 l_i<M, \text{for all}~i~~{\text{and}}~~|k_i|\geq \frac{N}{2} ~~\text{for~some}~i.
 \]

%     \begin{lemma}
% For multiple paths with the same aliased Doppler shift, the probability that more than one distinct delay has multiple overlapping paths is
% \[P_r=\]
% which is negligible for the considered system.

%  \textit{Proof:} See Appendix.
% \end{lemma}

At the transmitter, a set of $\mathcal Q$-QAM modulated information symbols, each with average energy $E_s$, are mapped onto the DD domain, denoted by $\mathbf X \in{\mathbb{C}^{M\times N}}$, which is then transformed to the time domain sample vector ${\mathbf s}\in \mathbb C^{MN\times 1}$ by an inverse discrete Zak transform (IDZT) as 
\cite{delay_Doppler_book}
 \[{\mathbf s}  \triangleq \text{IDZT}(\mathbf X)= \text{vec} (\mathbf X \mathbf F_N^H )=\text{vec}(\widetilde{\mathbf X})\in{\mathbb{C}^{MN\times 1}}, 
 \]
 where
\begin{equation}\label{EQ:IDZT}
    \widetilde{\mathbf X}=[\tilde{\mathbf x}_0^T,\ldots,\tilde{\mathbf x}_{M-1}^T]^T= \mathbf X \mathbf F_N^H \in{\mathbb{C}}^{M\times N}
\end{equation} 
represents the delay-time (DT) sample matrix with $\tilde{\mathbf x}_m\in{\mathbb{C}}^{N\times 1}$, $m\in[0, M-1]$ as the DT sample vector.

At the receiver, the time domain sample vector $\mathbf r \in{\mathbb{C}}^{MN\times 1}$ is converted to the DD domain using DZT as
 \[{\mathbf Y} \triangleq \text{DZT}(\mathbf r) = \text{vec}^{-1}(\mathbf r) \mathbf F_N=\widetilde{\mathbf Y}\mathbf F_N\in{\mathbb{C}^{M\times N}}, 
 \]
 where
\begin{equation}\label{EQ:IDZT}
    \widetilde{\mathbf Y}=[\tilde{\mathbf y}_0^T,\ldots,\tilde{\mathbf y}_{M-1}^T]^T= \text{vec}^{-1}(\mathbf r)\in{\mathbb{C}}^{M\times N}
\end{equation} 
represents the received DT sample matrix with $\tilde{\mathbf y}_m\in{\mathbb{C}}^{N\times 1}$, $m\in[0, M-1]$ as the received DT sample vectors.   Following \cite{delay_Doppler_book}, the received DD sample at $m$-th delay and $n$-th Doppler index can be expressed as
 \begin{align} 
\textbf Y[m,n] {\approx} \sum_{i=0}^{L-1} 
h_i e^{j\frac{2\pi}{MN} \left(m-l_{i}\right)k_i} \xi_i[m,n]\,
\textbf X\big[ [m{-}l_{i}]_M , [n{-}k_{i}]_N \big],
\end{align}

\begin{equation}\label{Eq:Xi}
\xi_i[m,n] =
\begin{cases}
1, & l_{i} \le m < M,\\
\dfrac{N-1}{N} e^{-j2\pi \left(\frac{[n-k_{i}]_N}{N}\right)}, & 0 \le m < l_{i}.
\end{cases}
\end{equation} 

\section{Proposed Channel Estimation for Overspread Doppler shifts}\label{Sec:PropCE}
This section describes the proposed training frame structure and the corresponding two-stage channel estimation for time-varying channels with overspread Doppler shifts.

 %DO YOU MEAN: the Doppler collision (different Dopplers coincide after modulo N) only happens in one delay tap?}. %(refer to Table II for the percentage occurrence of such cases).
 %\end{itemize}
\subsection{Proposed Training Frame}
Consider a discrete-time cosine pilot signal \[\mathbf c[m]=A~ \text{cos}\Big(2\pi \frac{f_0}{M} m\Big), ~m=0,\ldots,M-1,\]
sampled at $T_s=\frac{1}{M\Delta f}$, with amplitude $A$ and normalized discrete frequency $f_0=\frac{f}{\Delta f}$ corresponding to the cosine frequency $f$. We choose $f_0<M$ $\in \mathbb Z$ to ensure integer number of cycles over $M$ samples, thereby avoiding spectral leakage. Note that $M>k_{\rm max}$ guarantees that the maximum overspread Doppler shift can be identified in the frequency-domain peak detection. The proposed DD training frame is formulated by placing the cosine pilot signal in the first block along with a single high power pilot, $x_p$ $(|x_p|^2\gg A^2)$, in the first grid  as
\begin{equation}\label{EQ:TrainFrameDD}
    \mathbf X_{\rm t}[m,n]=\begin{cases} \mathbf c[0]+x_p &~~m=0, n=0, \\[1ex] 
    \mathbf c[m] &~~m\in[1,M-1], n=0,\\[1ex]
    0&~~\text{otherwise}, \end{cases}
\end{equation}
for $n\in [0,N-1]$, resulting in the time-domain training signal vector  
 \begin{equation}\label{td_OTFS}
    \mathbf s_{\rm t}\triangleq \text{IDZT}(\mathbf X_{\rm t})=\text{vec}\Big(\mathbf X_{\rm t} \mathbf F_N^H\Big)\in{\mathbb{C}^{MN\times 1}}. %= \text{vec}(\widetilde{\mathbf X}_{\text{t}}),
\end{equation}
The corresponding time domain received training signal $\mathbf r_{\rm t}[q]$, for $q\in[0,MN-1]$ is 
\begin{align}\label{td_rx_OTFS}
\mathbf r_{\rm t}[q]&=\sum_{i=0}^{L-1} \mathbf g[l_i,q]\mathbf s_{\rm t}[q-l_i] \notag\\
&= \sum_{i=0}^{L-1} h_ie^{\jmath\frac{2\pi}{MN}k_i(q-l_i)} \mathbf s_{\rm t}[q-l_i]+\mathbf w[q],
\end{align}
where 
\begin{align}\label{EQ:time-varying taps}
\mathbf g[l_i,q]=h_ie^{\jmath\frac{2\pi}{MN}k_i(q-l_i)},~~~\forall i
\end{align}
are the time-varying channel taps, and $\mathbf w[q]$ are the additive white Gaussian noise (AWGN) samples with zero mean and variance $\sigma^2_w$. Further, the DD received training matrix $\mathbf Y_{\rm t}\in{\mathbb{C}^{M\times N}}$ is obtained as \cite{delay_Doppler_book}
\begin{equation}\label{DDTraingMatrix}
    \mathbf Y_{\rm t}\triangleq \text{DZT}(\mathbf r_{\rm t})=\text {vec}^{-1}(\mathbf r_{\rm t})~\mathbf F_N.
\end{equation}
 
Next, we present the expression of the time domain received training signal assuming only the cosine pilot signal is transmitted. It will be utilized later for the overspread channel estimation. For such case, the element of DT transmit sample matrix $\tilde {\mathbf{X}}_{\rm c}$ is given by
\begin{equation}
   \tilde {\mathbf{X}}_{\rm c}[m,n]=\frac{\mathbf c[m]}{\sqrt N}=\frac{A}{\sqrt N}~\text{cos}\left(2\pi \frac{f_0}{M} m\right),
\end{equation}
resulting in the time-domain transmit sample vector as
\begin{equation}
   {\mathbf{s}_{\rm c}}[q]= \frac{A}{\sqrt N}~\text{cos}\left(2\pi \frac{f_0}{M} (q-nM)\right),
\end{equation}
for $ q\in[0,MN-1]$. Hence, the received signal is
 \begin{multline}\label{rxCosine}
    {\mathbf r}_{\rm c}[q]= \frac{A}{\sqrt N}
    \sum_{i=0}^{L-1}h_i e^{\jmath\frac{2 \pi}{MN}k_i(q-l_i)} \text{cos}\left(2\pi \frac{f_0}{M} (q-nM-l_i)\right)\\
    +\mathbf w[q].
\end{multline}

After applying Euler's formula to express the cosine function in its exponential form along with some mathematical manipulation, we can rewrite (\ref{rxCosine}) as
\begin{align}
    {\mathbf r}_{\rm c}[q]&= \frac{A}{2\sqrt N}\sum_{i=0}^{L-1}h_i e^{\jmath\frac{2 \pi}{MN}k_i(q-l_i)}\times\\ \nonumber
    & \left(e^{\jmath2\pi \frac{f_0}{M} (q-l_i)}e^{-\jmath2\pi {f_0} n}+e^{-\jmath2\pi \frac{f_0}{M} (q-l_i)}e^{\jmath2\pi {f_0} n}\right)+\mathbf w[q].
\end{align}
Since $f_0$ and $n$ are integers, $e^{\jmath2\pi {f_0} n}=e^{-\jmath2\pi {f_0} n}=1$ for all $n$.
Therefore,
\begin{align}\label{rXCosineTD}
    {\mathbf r}_{\rm c}[q]&= \frac{A}{2\sqrt N}\sum_{i=0}^{L-1}h_i\Big(e^{\jmath2 \pi(q-l_i)\left(\frac{k_i}{MN}+\frac{f_0}{M}\right)}\\  \nonumber
    & +e^{\jmath{2 \pi}(q-l_i)\left(\frac{k_i}{MN}-\frac{f_0}{M}\right)}\Big)+\mathbf w[q].
\end{align}

\subsection{Proposed Two-Stage Channel Estimation}
\begin{figure} 
\begin{center}
\setlength{\arrayrulewidth}{0.3mm}
\setlength{\tabcolsep}{5pt}
\renewcommand{\arraystretch}{1.3}
\begin{tabular}{|p{3.5mm}||p{7mm}|p{9mm}|p{9mm}|p{9mm}|p{7mm}|p{7mm}|} 
\hline
 $_l\backslash^{\kappa}$&0&1&2&3&4&5 \\
\hline \hline
0& {\sf X} $_{0}$ & & & & &   \\
\hline
1& & &  & & {\sf O} $_{1}$&   \\
 \hline
2& & {\sf O} $_{2}$& &  & &   \\
 \hline
3& & & & & &   \\
 \hline
4& & &{\sf O\hspace{-2.5mm}X} $_{3,4}$ & & &   \\
 \hline
5& & &{\sf O} $_{5}$  & & &   \\
 \hline
6& & {\sf O} $_{6}$ & & & &   \\
 \hline
7& & & &{\sf O\hspace{-3.5mm}\sf O} $_{7,8}$& &    \\
 \hline
\end{tabular}
\end{center}
\caption{DD received pilot echos with $M=8,N=6$ with $9$ paths $(l_i,k_i)=\{(0,0),(1,10), (2,13),(4,2),(4,14)$, $(5,20), (6,19), (7,3), (7,9)\}$. The path index  $i=0,\ldots, 8$ is given in each cell,  
{\sf X} represents an underspread path  ($k_i<\frac{N}{2}$) and {\sf O} an overspread path ($k_i\geq \frac{N}{2}$). 
} \label{over_ch_dd}
\end{figure}
%%%%%%%%%%%%%%%%%%%%%%%%%%%%%%%%%%%%
   \begin{figure}[t]
	 	\centering		\includegraphics[width=.9\linewidth]{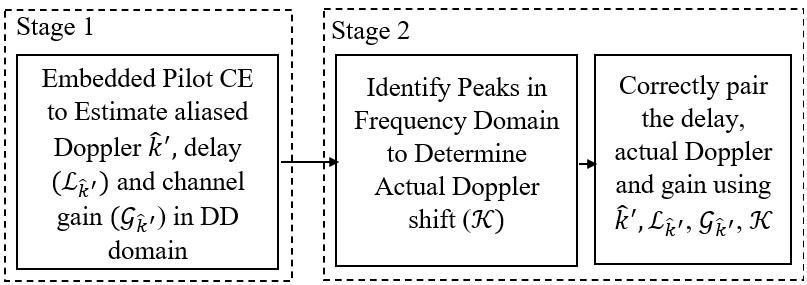}
	 	\caption{\small Flow Chart of Two-Stage CE}
        \label{flow_ch2stage}
  %\vspace{-.5cm}
	 \end{figure}
%%%%%%%%%%%%%%%%%%%%%%%%%%%%%%%%%%%%
  Let 
 \begin{equation}
     k'_i = \left\{ \begin{array}{ll}
 k_i & |k_i|< \frac{N}{2} \\
 \left[k_i\right]_N & \text{otherwise}
 \end{array}
 \right.
 \end{equation}
which can be observed  as the Doppler shift in the DD domain, and we will refer $k'_i$ as the {\em aliased Doppler shift throughout the paper}, since $|k'_i|<\frac{N}{2}$. %
 
 Fig. \ref{over_ch_dd} illustrates the received pilot echoes in the DD domain for the following cases. Note that these cases also apply for more than two paths. % to the group of paths with a group size greater than 2.}. 
\begin{itemize}
    \item {\em Case 1}. Paths have the same delay, with coinciding aliased Doppler shifts, e.g., paths $\{3,4\}$ and $\{7,8\}$.
    \item {\em Case 2}. Paths have different delays, with  coinciding aliased Doppler shifts, e.g., paths $\{3,5\}$, $\{2,6\}$, and $\{4,5\}$.
    %\item \sout{{\em Case 3}. Paths have different delays, with the same aliased Doppler shifts, e.g., $\{2,6\}$ and $\{4,5\}$.}
    \item {\em Case 3}. Paths have distinct delays and distinct aliased Doppler shift, e.g., path $0$ and $1$, respectively.
\end{itemize}
The traditional embedded pilot based CE can only handle the unspread Doppler shifts (e.g., path 0). %\sout{all the above cases except Case 1, where paths have same delays with coinciding normalized and aliased Doppler shifts.} 
 To tackle all the other cases, we propose a two-stage channel estimation method, which is illustrated in the flowchart in Fig. \ref{flow_ch2stage} and its pseudocode is provided in Algorithm~1. 

 Let $\hat{k}'_i$ be the estimated aliased Doppler shift. In the first stage,  an embedded pilot based CE is utilized to estimate the aliased Doppler shift $\hat{k}'_i $, delay $\hat l_i$, and channel gain $\hat h_i$ of the $i$-th path in the DD domain.

In the second stage, we transform the received time-domain cosine pilot signal into the frequency domain. Since the peaks of the cosine pilot signal appear at the actual Doppler shifts in the frequency domain, we are able to refine the Doppler estimate and accurately pair it with its corresponding delay and path gain using the estimates obtained in the first stage.

 \begin{algorithm}[h]
    		\caption{Two-stage CE }
    		\begin{algorithmic} [1]
    			%\STATE {Obtain
        % $\hat l_{\hat {\ell}_ij}=\hat {\ell}_i+b_\betaM, ~~\forall b_j\in \mathcal B$}
     \STATE {\textbf{Input:} $\mathbf r_{\rm t}, \mathbf s_{\rm t}, \mathbf Y_{\rm t}, \alpha, \sigma_w^2$}
      %  \STATE {$\mathbf{Input:} :~{\text{from first step CE}}: {\mathcal J},  \mathcal K_{\hat \ell_\mu} ~\forall \mu, \mathcal H$}.
      \STATE{\textbf{Output:} ${\mathcal H} =\{(\hat l_j, \hat k_j, \hat h_j)\}$}
        \STATE {{\bf Initialize:} Path list $\mathcal H=\{\}$, set ${\cal A}=\{\}, \iota=1$}

        \STATE{$\%$ ---- Stage 1 CE ----}
        \FOR {$n\rightarrow 0~\text{to}~N-1$}
        \STATE{Compute $\mathbf P[n]$ using (\ref{avgPowerDoppler})} 
%\IF{$\mathbf P[n]> \mathcal T_1=\alpha_1 \sigma_w^2$}
            \FOR {$m\rightarrow 0~\text{to}~M-1$}
        \IF{$|\mathbf Y_{\rm t}[m,n]|^2>A+ \alpha\mathbf P[n]$}
          \STATE{$\mathcal L_{\hat{k}'_\iota} =\{\}$, $\mathcal G_{\hat{k}'_\iota} =\{\}$}
        \STATE{$\lambda\leftarrow 1$}
           \STATE{$\hat{k}'_\iota \leftarrow n$, ${\cal A} \leftarrow  {\cal A} \cup \{{\hat{k}'_\iota}\}$ }
           \STATE{$ \hat l_{\lambda}({\hat{k}'_\iota})\leftarrow m$, 
        ${\mathcal L_{\hat{k}'_\iota}} \leftarrow  {\mathcal L_{\hat{k}'_\iota}} \cup \{ \hat l_{\lambda}({\hat{k}'_\iota})\}$ }
        \STATE{ Compute $ \hat h_{\lambda}({\hat{k}'_\iota})$ using (\ref{ComputegainDD}) }
        \STATE{$\mathcal G_{\hat{k}'_\iota}\leftarrow  {\mathcal G_{\hat{k}'_\iota}} \cup \{ \hat h_{\lambda}({\hat{k}'_\iota})\}$}
          \STATE{$\lambda\leftarrow \lambda+1$}
        \ENDIF
        \ENDFOR
          \STATE{$\iota\leftarrow \iota+1$}
 %       \ENDIF
        \ENDFOR
\STATE{$\%$ ---- Stage 2 CE ----}
 \STATE{$\text{\bf func}~[{\mathcal H}]=\textit{CosineFreqPeak} (\mathbf r_{\rm t}, \gamma_1, \gamma_2, \delta, x_p, \sigma_w^2, f_0,M,  N,$ \\
    ~~~~~~~~~~~~~~~~~~~~~~~~~~~~~~~~~~~~~~~~~~~~~~~~~~~~~~~~~~~~$ \mathcal H,\mathcal A,\mathcal L_{\hat{k}'_{\iota}},  \mathcal G_{\hat{k}'_{\iota}}, \forall \iota$)}
      \STATE{Return $\mathcal H$}
    		\end{algorithmic} \label{Alg_first_stage}
    	\end{algorithm}

        %%%%%%%%%%%%%%%%%%%%%%%%%%%%%%%%
 %\begin{algorithm}[h]
 
\subsubsection{First-Stage}

The training frame arrangement in (\ref{EQ:TrainFrameDD}) results in the received signal matrix in the DD domain %in (\ref{DDTraingMatrix}), 
containing both pilot and cosine pilot signal echoes, and can be further expressed as in (\ref{DDTrainFrame})\footnote{Here multiple paths with identical delays and aliased Doppler shifts are ignored, as they are indistinguishable in DD domain and are resolved in the subsequent estimation stage.}, where $\xi_i[m,n]$ in (\ref{Eq:Xi}) for $n=k'_i=[k_i]_N$ reduces to 
\[\xi_i[m,n]=\begin{cases} 1, &~~ l_i\leq m<M\\[1ex]
\frac{N-1}{N}&~~0\leq m<l_i \end{cases}.\]

% \begin{figure*}
% \begin{equation}\label{DDTrainFrame}
%     \mathbf Y_{\rm t}[m,n]=\begin{cases}h_i x_p+\sum_{i^{\prime}=0, i^{\prime}\neq i}^{L-1} h_{i^{\prime}} e^{\jmath\frac{2\pi}{MN} k_{i^{\prime}} (m-l_{i^{\prime}})} \xi_{i^{\prime}}[m,n]~\text{cos}(2 \pi \frac{f_0}{M}[m-l_{i{^{\prime}}}]_M)+\mathbf w[m,n] &~~ m=l_i,n=k'_i=[k_i]_N\\[1ex] 
%     \sum_{i=0}^{L-1} h_{i} e^{\jmath\frac{2\pi}{MN} k_i (m-l_i)} \xi_i[m,n]\text{cos}(2 \pi \frac{f_0}{M}[m-l_i]_M)+\mathbf w[m,n] &~~ m\neq l_i,n=k'_i=[k_i]_N\\[1ex] 
%     \mathbf w[m,n], &~~~ otherwise
%     \end{cases}
% \end{equation}
% \end{figure*}
\begin{figure*}
\begin{equation}\label{DDTrainFrame}
    \mathbf Y_{\rm t}[m,n]=\begin{cases}h_i (A+x_p)+\mathbf w[m,n] &~~ m=l_i,n=k'_i=[k_i]_N
    \\[1ex] 
    A h_{i} e^{\jmath\frac{2\pi}{MN} k_i (m-l_i)} \xi_i[m,n]\text{cos}(2 \pi \frac{f_0}{M}[m-l_i]_M)+\mathbf w[m,n] &~~ m\neq l_i,n=k'_i=[k_i]_N\\[1ex] 
    \mathbf w[m,n], &~~~ {\text{otherwise}}
    \end{cases}
\end{equation}
\end{figure*}

Based on (\ref{DDTrainFrame}), the received signal exhibits pilot echoes corresponding to the delay and aliased Doppler shift values.  Further, the cosine pilot signal $\{\mathbf c[m]\}_{m=0}^{M-1}$  is echoed at all the delay taps with indices $m=[0,M-1]$ corresponding to the aliased Doppler shift indices $k'_i$. The remaining received DD signal samples contain only noise. Hence, to identify the echoed pilot locations, we employ an adaptive threshold
\[\mathcal T_{n}=A+\alpha \mathbf P[n], ~~\alpha>0\]where 
 \begin{equation}\label{avgPowerDoppler}
     \mathbf P[n]=\frac{1}{M}\sum_{m=0}^{M-1}|\mathbf Y_{\rm t}[m,n]|^2.
 \end{equation}
 The adaptive threshold $\mathcal T_n$ eliminates all DD grid points that contain only noise, as well as those that contain the cosine pilot signal superimposed with noise.
 
   We denote the initial sets of estimated aliased Doppler shifts $\hat{k}'_\iota$, $\forall{\iota}$, and the associated estimated path gains and delays as 
$ \mathcal A,~~\mathcal G_{{\hat{k}'_\iota}},~~\text{and}~~\mathcal L_{{\hat{k}'_\iota}} $, respectively.

 We obtain the estimates of the aliased Doppler shifts $\hat{k}'_\iota$, and all the corresponding delays $ \hat l_{\lambda}(\hat{k}'_\iota)$ and the path gains corresponding to $\{\hat l_{\lambda}(\hat{k}'_{\iota}), \hat{k}'_{\iota}\}$, where $\lambda$ denotes the index of the estimated delays/corresponding path gain estimates, as
\begin{align}\label{ComputegainDD}
    &\hat{k}'_{\iota} =n,~\hat l_{\lambda}(\hat{k}'_\iota)=m, ~~ \text{if} ~~|\mathbf Y_{\rm t}[m,n]|^2> \mathcal T_{n},\notag\\
    &\hat h_{\lambda}(\hat{k}'_{\iota})=\mathbf Y_{\rm t}[\hat l_{\lambda}(\hat{k}'_{\iota}), \hat{k}'_{\iota}]/(A+x_p).
\end{align}
%and the path gain corresponding to the delay-Doppler pair $\{\hat l_{\lambda}(\hat{k}'_{\iota}), \hat{k}'_{\iota}\}$ as
% % \begin{equation}\label{ComputegainDD}
%      \hat h_{\lambda}(\hat{k}'_{\iota})=\mathbf Y_{\rm t}[\hat l_{\lambda}(\hat{k}'_{\iota}), \hat{k}'_{\iota}]/(A+x_p).
%  \end{equation}
and then place these estimates into the sets 
\[
 \mathcal A=\{\hat{k}'_\iota\},~~\mathcal L_{{\hat{k}'_\iota}}=\{\hat l_{\lambda}({\hat{k}'_\iota})\}~\text{and}~ \mathcal G_{{\hat{k}'_\iota}}=\{\hat h_{\lambda}({\hat{k}'_\iota})\},
 \]
 where $|\mathcal L_{{\hat{k}'_\iota}}|=|{\cal G}_{{\hat{k}'_\iota}}|$. %{\color{red} note sure this is needed}.% is the index of the set\footnote{ Note that $ |\mathcal L_{{\hat{k}'_\iota}}|=|\mathcal G_{{\hat{k}'_\iota}}|$}.  

 When multiple paths have distinct delays but with the same aliased Doppler shifts (see Case 2, paths $\{2,6\}$ and $\{3,5\}$), the estimation yields $|\mathcal L_{{\hat{k}'_{\iota}}}|>1$; otherwise, when paths has same delay but with coinciding aliased Doppler shifts (see Case 1, paths $\{7,8\}$ or $\{3,4\}$), the estimation can only identify one path with incorrect path gain, i.e., $|\mathcal L_{{\hat{k}'_{\iota}}}|=1$. Furthermore, when paths have distinct delays and distinct aliased Doppler shifts (see Case 3, paths $0$ and $1$), the estimation for each path provides the distinct delay estimate, i.e., $|\mathcal L_{{\hat{k}'_{\iota}}}|=1$. %{\color{red} missing case 2}

% \noindent \textit{Example 2:} Consider the example in Fig. \ref{over_ch_dd}. Here, $\mathcal A=\{0, 1, 2, 3, 4\}$, $\mathcal L_0=\{0\}$, $\mathcal L_1=\{2,6\}$, $\mathcal L_2=\{4,5\}$, $\mathcal L_3=\{7\}$, $\mathcal L_4=\{1\}$. 

% Paths 2, 6 have same aliased Doppler $1$ but distinct delay. Hence $|\mathcal L_1=\{2,6\}|>1$. Paths 7, 8 have same aliased Doppler $3$ and same delay, hence, we estimate one path with incorrect channel gain and $|\mathcal L_3=\{7\}|=1$. Similarly, between paths 3 and 4, we estimate only one path. For the paths (0 and 1) with distinct Doppler, i.e., $\{0, 4\}$, we have $|\mathcal L_0=\{0\}|=|\mathcal L_4=\{1\}|=1$.

\subsubsection{Second Stage CE in Frequency Domain}
To estimate the actual Doppler shift, we transform the received time domain signal into the frequency domain to obtain peaks at the Doppler frequencies caused by the  cosine pilot signal echoes. For this, we first remove the high power pilot samples from the received training vector $\mathbf r_{\rm t}$ to isolate the received cosine pilot signal $\mathbf r_{\rm c}$. This is done by inserting zeros at the identified delay locations in the DT received training frame as
\begin{equation}\label{rxDtcosine}
\tilde {\mathbf Y}_{\rm c}[m,n]=\begin{cases}0 & m=\hat l_{\lambda}(\hat{k}'_{\iota}), n\in\{0,N-1\}\\[1ex]\tilde{\mathbf Y}_{\rm t}[m,n] & \text{otherwise} \end{cases}
\end{equation}
where $\tilde{\mathbf Y}_{\rm t}=\mathbf Y_{\rm t} {\bf F}_N^{-1}$ denotes the DT domain received training matrix prior to the above zero insertion.
The received cosine vector is then obtained as
\begin{equation}\label{rxcheckcosine}
     \check{\mathbf r}_{\rm c}=\text{vec}(\tilde {\mathbf Y}_{\rm c})
\end{equation}
The resulting time domain received cosine pilot signal is then converted to the frequency domain by using an $MN$-point DFT as
\begin{equation}\label{FDCosine}
    \mathbf R[f]=\frac{1}{\sqrt{MN}}\sum_{q=0}^{MN-1} \check{\mathbf r}_{\rm c}[q]e^{-\jmath\frac{2 \pi}{MN}fq}\\
\end{equation}
Using (\ref{rXCosineTD}) in (\ref{FDCosine}) and rearranging the terms, we obtain
\begin{align}
    \mathbf R[f]&=\frac{A}2N\sqrt{ M}\sum_{i=0}^{L-1}h_i \times \\\nonumber
    &\Big(e^{-\jmath2 \pi\left(\frac{k_i}{MN}+\frac{f_0}{M}\right)l_i}\underbrace{\sum_{q=0}^{MN-1} e^{\jmath2 \pi\left(\frac{k_i}{MN}+\frac{f_0}{M}-\frac{f}{MN}\right)q}}_{(a)} +\\  \nonumber
    &e^{-\jmath2 \pi\left(\frac{k_i}{MN}-\frac{f_0}{M}\right)l_i}\underbrace{\sum_{q=0}^{MN-1} e^{\jmath2 \pi\left(\frac{k_i}{MN}-\frac{f_0}{M}-\frac{f}{MN}\right)q}}_{(b)} \Big)+\mathbf W[f] \\  \nonumber
    \end{align}
    where $f=\frac{-MN}{2}+1, \ldots, \frac{MN}{2}$. The terms $(a)$ and $(b)$ equal to $MN$ for $\left(\frac{k_i}{MN}+\frac{f_0}{M}-\frac{f}{MN}\right)=0$ and $\left(\frac{k_i}{MN}-\frac{f_0}{M}-\frac{f}{MN}\right)=0$, respectively; otherwise, both terms become zero. Hence, 
    \begin{equation}\label{rxFDcosine}
         \mathbf R[f]=\begin{cases} \frac{ A \sqrt{M}}{2} h_i e^{-\jmath\frac{2 \pi}{MN}fl_i}+\mathbf W[f], &~~ f=k_i+f_0N\\
    \frac{ A \sqrt{M}}{2} h_i e^{-\jmath\frac{2 \pi}{MN}fl_i}+\mathbf W[f], &~~ f=k_i-f_0N\\
    \mathbf W[f], &~~~~ \text{otherwise}.\end{cases}
    \end{equation}

Based on (\ref{rxFDcosine}), the actual Doppler shift can be computed as
\begin{equation}\label{estimateActualDopler}
     \hat k_i=f-f_0N, ~~ \text{if} ~~|\mathbf R[f]|^2>  \gamma \sigma_w^2, ~~\gamma>0,
\end{equation}
for $f=0, \ldots, \frac{MN}{2}$. Note that Doppler shifts identified in the range $~f=[-\frac{MN}{2}+1, -1]$ are aliased counterparts of $~f=[0,\frac{MN}{2}]$, as inferred from (\ref{rxFDcosine}). Hence, it can be ignored.

Let $\mathcal K=\{\hat k_i,~i=0,\ldots, L-1\}$ be the set containing all the actual Doppler shifts based on (\ref{estimateActualDopler}). 
Next, we are required to {\em pair the estimated actual Doppler shifts with their corresponding delays and channel gains} using the information $\mathcal A, \mathcal K,  \mathcal L({\hat{k}'_\iota}),$ and $\mathcal G({\hat{k}'_\iota}), \forall \iota$ by following steps. Note that all these steps are summarized as the second stage channel estimation algorithm in Appendix.

\begin{itemize}
    \item \textbf {Step 1:} Determine the aliased Doppler shifts corresponding to the estimated actual Doppler shifts in $\mathcal K$ to form the set
   \begin{equation}\label{eq:aliasedDoppset}
       \mathbb K=\left\{\mathbbm k_{i}=[\hat k_{i}]_N , \text{for~all}~\hat k_{i}\in \mathcal K\right\} 
    \end{equation}
     We then denote the set $\tilde{\mathbb K}=\{\mathbbm k_{\tilde i}\}$ as the set containing the unique elements of $\mathbb K$.
    \item \textbf{Step 2:} For each $\tilde i = 1,\ldots,|\tilde {\mathbb K}|$, we define the lists of Dopplers which overlap with ${\mathbbm k}_{\tilde i}$ when aliased as %let $\mathbb K_{\tilde i}=\{\}$. We append all the actual Doppler shifts in the set $\mathcal K$ to $\mathbb K_{\tilde i}$ as
   \[\mathbb K_{\tilde i}=\{\hat k_i|[\hat k_i]_N={\mathbbm k}_{\tilde i}\}\]
   
    Let the elements in $\mathbb K_{\tilde i}$ be denoted by $\hat k_{\beta}(\mathbbm k_{\tilde i})$, where $\beta= 1,\dots,|\mathbb K_{\tilde i}|$. Then for each $\mathbbm k_{\tilde i}$, we find $\hat{k}'_\iota\in \mathcal A$ such that $\hat{k}'_\iota=\mathbbm k_{\tilde i}$. 
    \item \textbf {Step 3:} If $|\mathbb K_{\tilde i}|=1$, the aliased Doppler shift $\mathbbm k_{\tilde i}={\hat{k}'_\iota}$ is unique (e.g. Case 3), and thus $|\mathcal L_{{\hat{k}'_\iota}}|=1$. 
     Hence, we estimate 
    \[\hat l_{j}=\hat l_{1}(\hat{k}'_\iota=\mathbbm k_{\tilde i}),~~ \hat k_{j}=\hat k_{1}(\mathbbm k_{\tilde i}),~~\hat h_{j}=\hat h_{1}(\hat{k}'_\iota)\] and include them into the channel estimate path list 
    \begin{equation}\label{UpdatePathlist}
         \mathcal H= \{\hat l_j,\hat k_j,\hat h_j\}.
     \end{equation}  
   %  \item \textbf {Step 4:} If $|\mathbb K_{\tilde i}|=|\mathcal L_{\hat{k}'_{\iota}}|>1$, i.e., multiple paths have same aliased Doppler shift but with distinct delays (e.g., Case 2), using (\ref{rxFDcosine}), we pair the delay, actual Doppler shift, and channel gain and obtain the estimates    
   %      \[\hat l_{j}=\hat l_{\lambda}(\hat{k}'_\iota=\mathbbm k_{\tilde i}),~~ \hat k_{j} = \hat k_{\beta}(\mathbbm k_{\tilde i}),~~ \hat h_{j} = \hat h_{\lambda}(\hat{k}'_\iota)\]
   %  such that they satisfy 
   %  \begin{align}\label{FreqDomCondition}
   %     & \big|\frac{2}{A\sqrt{M}}\mathbf R({\hat k}_{\beta}(\mathbbm k_{\tilde i}) +f_0N)\big|- |\hat h_{\lambda}(\hat{k}'_\iota)|\notag\\
   %     % & = |h_i - \hat h_{\lambda}(\hat{k}'_\iota)|<\delta,\\ \nonumber
   %        & = |h_i| -| \hat h_{\lambda}(\hat{k}'_\iota)|<\epsilon',
   %  \end{align} 
   % for $\beta= 1,\dots,|\mathbb K_{\tilde i}|$, and $\epsilon'>0$ is a small threshold value.   This condition selects the delay-Doppler and channel gain combination such that the magnitude of the reconstructed channel gain, obtained using (22), is closest to the magnitude of the first-stage channel gain estimate $\hat h_{\lambda}(\hat{k}'_\iota)$ within a small tolerance $\epsilon'$.} We include the estimates to the path list $\mathcal H$ as in (\ref{UpdatePathlist}).
       \item   \textbf {Step 4:} If $|\mathbb K_{\tilde i}|=|\mathcal L_{\hat{k}'_{\iota}}|>1$, i.e., multiple paths have same aliased Doppler shift but with distinct delays (e.g., Case 2), then for each delay, the actual Doppler shift from the candidate Doppler set $|\mathbb K_{\tilde i}|$ is selected by
minimizing the absolute difference between the magnitude of the reconstructed channel gain obtained using (\ref{rxFDcosine}) and the magnitude of the first-stage channel gain estimate corresponding to that delay i.e., we select the Doppler index as
        \begin{align}\label{FreqDomCondition}
       &\beta^\star(\lambda)=\argmin\limits_{\beta} \big|\big|\frac{2}{A\sqrt{M}}\mathbf R({\hat k}_{\beta}(\mathbbm k_{\tilde i}) +f_0N)\big|- |\hat h_{\lambda}(\hat{k}'_\iota)|\big|\notag\\
       % & = |h_i - \hat h_{\lambda}(\hat{k}'_\iota)|<\delta,\\ \nonumber
          & = \argmin\limits_{\beta}\big||\hat h_i| -| \hat h_{\lambda}(\hat{k}'_\iota)|\big|,
    \end{align} 
       
        for $\beta= 1,\dots,|\mathbb K_{\tilde i}|$ and obtain the estimates    
        \[\hat l_{j}=\hat l_{\lambda}(\hat{k}'_\iota=\mathbbm k_{\tilde i}),~~ \hat k_{j} = \hat k_{\beta^*(\lambda)}(\mathbbm k_{\tilde i}),~~ \hat h_{j} = \hat h_{\lambda}(\hat{k}'_\iota)\]
   
  This condition selects the delay-Doppler and channel gain combination such that the magnitude of the reconstructed channel gain, obtained using (\ref{rxFDcosine}), is closest to the magnitude of the first-stage channel gain estimate $\hat h_{\lambda}(\hat{k}'_\iota)$ corresponding to the delay. We include the estimates to the path list $\mathcal H$ as in (\ref{UpdatePathlist}).
     \item \textbf {Step 5:} For $|\mathbb K_{\tilde i}|>1$ and $|\mathcal L_{\hat{k}'_{\iota}}|=1$, i.e., multiple paths have the same aliased Doppler shift and the same delay (e.g., Case 1, Paths $\{7,8\}$) %we assign the delay $\hat l_{1}(\hat{k}'_\iota)$ to all the Dopplers in $\mathbb K_{\tilde i}$ to estimate
     we estimate 
     \[\hat l_{j}=\hat l_{1}(\hat{k}'_\iota=\mathbbm k_{\tilde i}), \hat k_{j}=\hat k_{\beta}(\mathbbm k_{\tilde i}).\]
     for $\beta= 1,\dots,|\mathbb K_{\tilde i}|$.
     Since {\em only one path with incorrect channel gain is estimated in the first stage}, we set temporarily the channel gains to be $\hat h_j=0, \forall \beta$. They will be  finely estimated in Step 7. %The path list is updated using (\ref{UpdatePathlist}).
     \item \textbf{Step 6:} For $|\mathbb K_{\tilde i}|> |\mathcal L_{\hat{k}'_{\iota}}|>1 $, i.e., among the paths with same aliased Dopplers, some paths have same delay while the other paths have distinct delays, (e.g., Case 2, Paths $\{3,4,5\}$). The paths with distinct delays satisfying (\ref{FreqDomCondition}) can be estimated using Step 4. The remaining paths with same delays can be estimated using Step 5. The path list is updated as in (\ref{UpdatePathlist}).
     \item \textbf{Step 7:} Finally, the channel gains for the paths that are set to $\hat h_j=0$ are estimated in the time domain using the least squares (LS) method. We first reformulate the time-domain input-output relation in vector form as % (\ref{td_rx_OTFS}) as
     \begin{equation} \label{EQ:LS}
         {\mathbf r}'_{\rm t}=\mathbf \Phi \hat {\mathbf h}
     \end{equation}
     where $\mathbf \Phi$ is $\tilde{Q}\times \tilde{L}$ matrix with entries $\mathbf \Phi[\tilde q,j]=e^{\jmath\frac{2 \pi}{MN}\hat {k}_{j }(\tilde q-{\hat l_{j})}}~\mathbf{ s}_{t} [\tilde q-\hat l_{j}]$, $\tilde Q <MN$ is the number of received signal samples used for the least square channel gain estimation, i.e., 
     \begin{align}
         {\mathbf r}'_{\rm t} &=  [{\mathbf r}'_{\rm t}[0], \ldots, {\mathbf r}'_{\rm t}[\tilde q], \ldots, {\mathbf r}'_{\rm t}[\tilde Q {-1}] ]^T\in \mathbb C^{\tilde Q\times 1},
    \end{align}
     and $\tilde L=|\mathcal H|$ is the total number of estimated paths with the estimated gains 
     \begin{align}
    \hat {\mathbf h} &=[\hat {\mathbf h}[0],\ldots, \hat{\mathbf h}[j],\ldots,\hat {\mathbf h}[\tilde L-1]]^T \in \mathbb C^{\tilde L\times 1}.
    \end{align}
   obtained from the LS solution of (\ref{EQ:LS})
     \begin{equation}\label{chann_Est_LS}
         \hat {\mathbf h}=(\mathbf \Phi^H\mathbf \Phi)^{-1}\mathbf \Phi^H  {\mathbf r}'_{\rm t}.
     \end{equation}
     The channel gains previously set to  $\hat h_j=0$ are replaced by the corresponding elements in $\hat {\mathbf h}[j]$.
    \end{itemize} 
\noindent \textit{Example 2:} In Fig. \ref{over_ch_dd} with $M=8$, $N=6$ and $9$ paths, Steps 1 and 2 yields 
\begin{align*}
    &\mathcal K=\{0, 10, 13, 2, 14, 20, 19, 3, 9\}, ~~\tilde{\mathbb K}=\{0, 1, 2, 3, 4\} \\
     &\mathbb K_1=\{0\}, ~~\mathbb K_2=\{13,19\} \\
    & \mathbb K_3=\{2,14,20\}, ~~\mathbb K_4=\{3,9\}, ~~\mathbb K_5=\{10\}
\end{align*}%\[\mathcal K=\{0, 10, 13, 2, 14, 20, 19, 3, 9\}, ~~\tilde{\mathbb K}=\{0, 1, 2, 3, 4\}\] ~~\[,\] ~~\[ \mathbb K_3=\{2,14,20\}, ~~\mathbb K_4=\{3,9\}, ~~\mathbb K_5=\{10\}\] 

$\mathbb K_1$ satisfies the condition in Step 3. Here, we have $\mathbbm k_1=\hat{k}'_1=0$ and $\mathcal L_0=\{0\}$. Hence, we find the  delay and Doppler tap estimates $(0,0)$ for the path 0.  Similarly, $\mathbb K_5$ also satisfies the condition in Step 3 and we find the  delay and Doppler estimates $(1,10)$ for the Path 1.
$\mathbb K_2$ satisfies the condition in Step 4. Here, we have $\mathbbm k_2=\hat{k}'_2=1$ and $\mathcal L_1=\{2,6\}$. Using the condition in (\ref{FreqDomCondition}), we can correctly pair $(2, 13)$ and $(6, 19)$ as the estimate for Paths 2 and 6, respectively.

$\mathbb K_4$ satisfies the condition in Step 5. We have $\mathbbm k_4=\hat{k}'_4=3$ and $\mathcal L_3=\{7\}$. Thus, we pair $(7, 3)$ and $(7, 9)$ as the delay and Doppler estimates for Paths 7 and 8, respectively and set their channel gains $\hat h_j=0$.

$\mathbb K_3$ satisfies the condition in Step 6. We have $\mathbbm k_3=\hat{k}'_3=2$ and $\mathcal L_2=\{4,5\}$. Path 5 satisfies the condition in (\ref{FreqDomCondition}) and hence we find delay and Doppler estimates $(5, 20)$  using step 4. The remaining Doppler shifts $2,14$ in $\mathbb K_3$ are paired with delay $4$, resulting in the delay and Doppler estimates $(4,2)$ and $(4,14)$, with their channel gains setting to be $\hat h_j=0$ as in Step 5.

Using (\ref{chann_Est_LS}), we estimate the gains of Paths 3, 4, 7, and 8, while the gains for the remaining paths were correctly estimated in the first stage.
 %%%%%%%%%%%%%%%%%%%%
 \subsection{Complexity}
     For the conventional embedded-pilot-based channel estimation, the dominant complexity arises from OTFS demodulation involving FFT processing, resulting in an overall complexity of\[\mathcal{O}(MN\log_2(N)).\]In the proposed scheme, additional computations are introduced in the two-stage channel estimation process. In the first stage, computation of the threshold $\mathcal{T}_n$ using (\ref{avgPowerDoppler}) requires $MN$ complex multiplications, resulting in complexity $\mathcal{O}(MN)$. Further, in the second-stage, the dominant operations include:
    (i) conversion of the received signal from time domain to frequency domain using an $MN$-point FFT with complexity\[\mathcal{O}(MN\log_2(MN)),\]and (ii) the LS-based estimation of channel gains in (\ref{chann_Est_LS}), which incurs complexity\[\mathcal{O}(\tilde{Q}\tilde{L}^2 + \tilde{L}^3).\]The LS step is required only in scenarios where multiple paths share the same delay and aliased Doppler shifts (e.g., Case~1), and hence does not affect all channel realizations. 
    
    Therefore, compared to the conventional embedded-pilot-based channel estimation, the proposed scheme introduces an additional complexity of\[\mathcal{O}\left(MN+MN\log_2(MN) + \tilde{Q}\tilde{L}^2 + \tilde{L}^3\right).\]

\subsection{PAPR and Spectral Efficiency}
 The peak-to-average power ratio (PAPR) of the proposed training signal is 
comparable to that of the standard embedded pilot based CE schemes, as both employ similar pilot arrangement. In our design, the training frame contains 
$MN - M$ zeros in the DD domain, and the cosine signal together with the single pilot 
symbol spreads across these zeros in the delay-time domain, resulting in a reduced 
PAPR. The computed PAPR value of the proposed training frame for $M=512, N=64$ and $\Delta \mathrm{SNR} = 28~\mathrm{dB}$ is 25.7 dB. Here $\Delta\text{SNR}=\text{SNR}_p-\text{SNR}_c$, where  $\text{SNR}_{\rm p}=\frac{|x_{\rm p}|^2}{N\sigma_w^2}$ and $\text{SNR}_{\rm c}=\frac{A^2}{N\sigma_w^2}$ denote {\em pilot} and {\em cosine pilot signal SNRs} in the time domain, respectively.  Under the same system configuration, the embedded pilot channel estimation \cite{embedded_pilot_raviteja} exhibits a PAPR of $25$ dB which is comparable to our proposed training frame. Here  $\Delta\text{SNR}=\text{SNR}_p-\text{SNR}_d$, where  $\text{SNR}_{\rm p}$ and $\text{SNR}_{\rm d}$ denote {\em pilot} and {\em data signal SNRs} in the time domain, respectively. 

 Regarding spectral efficiency, one OTFS frame is dedicated for channel estimation, followed by the transmission of multiple data frames. Hence, the spectral efficiency of the proposed scheme depends on the transmission periodicity of the training frame, which is determined by the \emph{channel geometric coherence time}, i.e., the duration over which the delay-Doppler paths remain approximately unchanged. For a typical urban terrestrial scenario, the geometric coherence time can be on the order of several tens of milliseconds. For example, at a vehicle speed of 100 kmph, the displacement over 50 ms is approximately 1.39 m, which typically does not significantly alter the channel path geometry and the associated normalized delay and Doppler values. For an OTFS frame with $N=64$ and $\Delta f=15$ KHz, the frame duration is 4.26 ms. Thus, one training frame followed by approximately 10 data frames i.e., nearly 11 OTFS frames can be transmitted before re-estimating the channel.

With the above-mentioned values, the pilot overhead of the proposed training framework is approximately \(9\%\), resulting in a spectral efficiency of approximately \(91\%\). In contrast, the embedded pilot-based channel estimation method requires pilot and guard resources of size \(l_{\max}\times 2 \times N\). In overspread Doppler scenarios, where the Doppler spread can exceed \(N/2\), guard symbols are required across the entire Doppler dimension. Considering a typical value of maximum normalized delay \(l_{\max}=20\), $M=512$, and assuming that the pilot and guard resources are utilized only in the first frame followed by \(10\) consecutive data frames, the pilot overhead and the resulting spectral efficiency is approximately $0.7\%$ and \(99.3\%\) respectively. 

Therefore, the embedded pilot-based channel estimation achieves higher spectral efficiency. However, it fails to estimate the overspread Doppler shifts entirely as demonstrated in the result section (see Figs. \ref{nmsedeltauniform} and \ref{berdeltauniform}), whereas the proposed training framework is specifically designed for reliable channel estimation in such scenarios. Hence, the proposed scheme trades spectral efficiency for improved channel estimation capability in overspread Doppler environments.
  %%%%%%%%%%%%%%%%%%%
   
\section{Results and Discussions}\label{Sec:Simulation}
\begin{table}[t]
\caption{\small Simulation Parameters for Scenarios A and B}
\begin{center}
\begin{tabular}{ | m{5em} | m{4.5em} | }
\hline
\vspace{0.1 cm}Parameter & \vspace{0.1 cm}Value \\[1ex]
\hline\hline

\vspace{0.1 cm}$\Delta f$ & \vspace{0.1 cm}15 KHz \\[1ex]
\hline

\vspace{0.1 cm}$M$ & \vspace{0.1 cm}$512$ \\[1ex]
\hline

\vspace{0.1 cm}$N$ & \vspace{0.1 cm}$64$ \\[1ex]
\hline

\vspace{0.1 cm}Modulation & \vspace{0.1 cm}4-QAM \\[1ex]
\hline

\vspace{0.1 cm}$f_0$ & \vspace{0.1 cm}50 \\[1ex]
\hline

\end{tabular}
\label{table_chann_param}
\end{center}
\end{table}
In this section, we present the performance of the proposed channel estimation in terms of normalized mean square error (NMSE) and bit error rate (BER). We define NMSE as
 
\[
\mathrm{NMSE}=\mathbb E\Big[\frac{\left\lVert \widehat{\mathbf g}-\mathbf g \right\rVert^2}{\left\lVert \mathbf g \right\rVert^2}\Big],
\]
where $\mathbf g \in \mathbb{C}^{\max(\mathcal{L} \cup \tilde{\mathcal{L}})\times MN}$ denotes the actual delay-time channel matrix, whose entries are defined as
\[
\mathbf{g}[m',q]=
\begin{cases}
\mathbf g[l_i,q], & \text{if } m' = l_i,\\
0, & \text{otherwise},
\end{cases}
\]
for $m' = [0,\max(\mathcal{L} \cup \tilde{\mathcal{L}})-1]$, where $\mathbf g[l_i,q]~\forall i$, is given in (\ref{EQ:time-varying taps}).

Similarly, $\widehat{\mathbf g} \in \mathbb{C}^{\max(\mathcal{L} \cup \tilde{\mathcal{L}})\times MN}$ denotes the estimated delay-time channel matrix, with entries
\[
\widehat{\mathbf{g}}[m',q]=
\begin{cases}
\widehat {\mathbf g}[\hat l_j,q], & \text{if } m' = \hat l_j,\\
0, & \text{otherwise},
\end{cases}
\]
where $\widehat {\mathbf g}[\hat l_j,q]$, for all $j$, is obtained using (\ref{EQ:time-varying taps}) with estimated values.

We present the performance of the proposed channel estimator for the two different scenarios: {\em Scenario A} and {\em Scenario B}. The common simulation parameters for both scenarios are summarized in Table~\ref{table_chann_param}.

 To evaluate the BER performance, we employ maximum ratio combining (MRC) detection for zero-pading OTFS (ZP-OTFS) \cite{tharaj_rake_MRC_journal} using a damping factor $\delta=0.25$ and maximum 10 iterations. We assume the training frame is transmitted prior to the data frame, during which the channel remains unchanged. We denote {\em pilot}, {\em cosine pilot signal}, and {\em data SNRs} in the time domain by $\text{SNR}_{\rm p}=\frac{|x_{\rm p}|^2}{N\sigma_w^2}$, $\text{SNR}_{\rm c}=\frac{A^2}{N\sigma_w^2}$, and $\text{SNR}_{\rm d}=\frac{E_s}{\sigma_w^2}$, respectively, and the difference between the pilot and cosine pilot signal SNRs as $\Delta\text{SNR}=\text{SNR}_p-\text{SNR}_c$. 

%%%%%%%%%%%%%%
\subsection{Scenario A}
       \begin{figure} 
	 	\centering		\includegraphics[width=.9\linewidth]{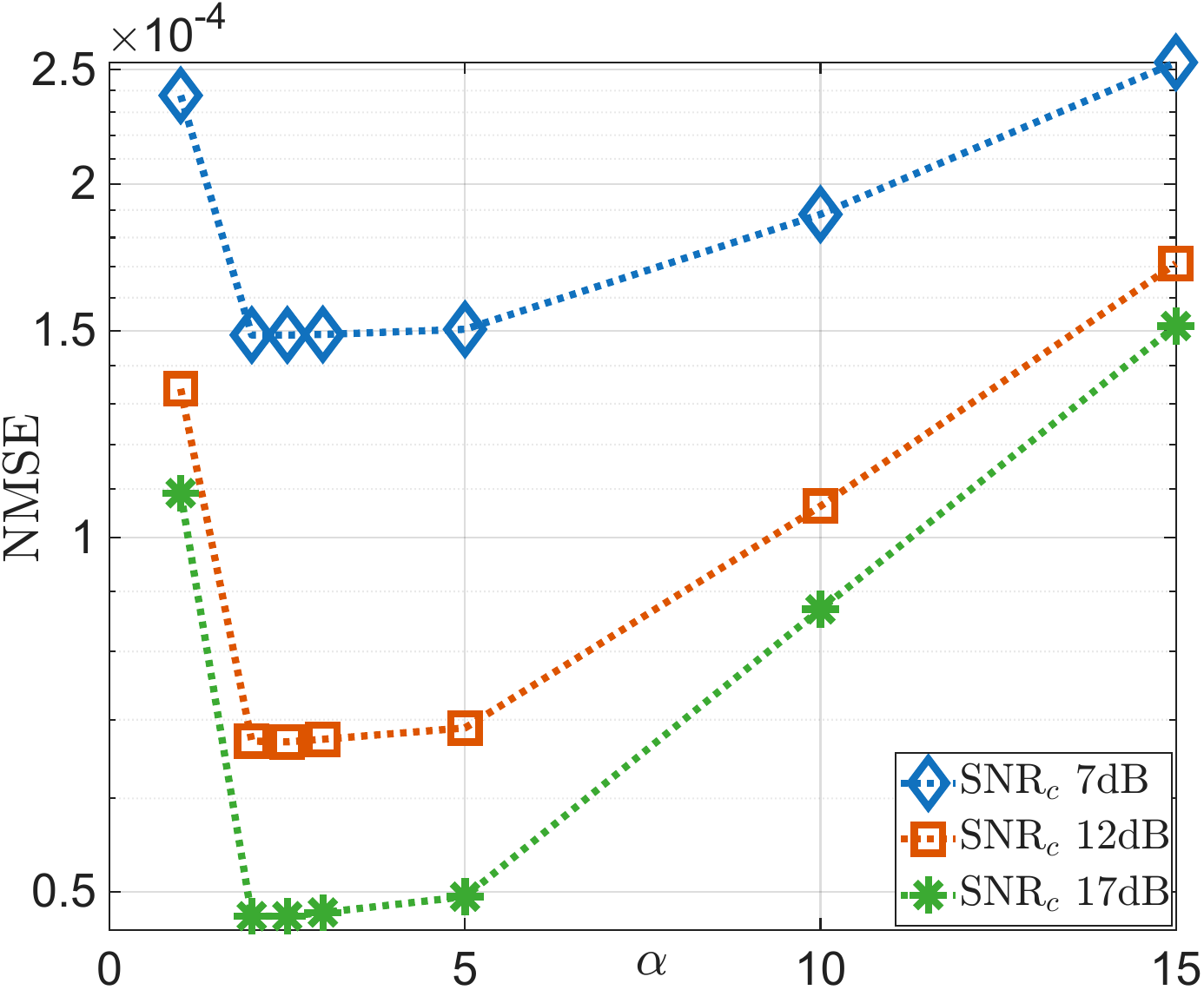}
	 	\caption{\small NMSE vs first stage threshold parameter $\alpha$ for Scenario A}
          \label{nmsealphauniform}
  %\vspace{-.5cm}
	 \end{figure}
       \begin{figure} 
	 	\centering		\includegraphics[width=.9\linewidth]{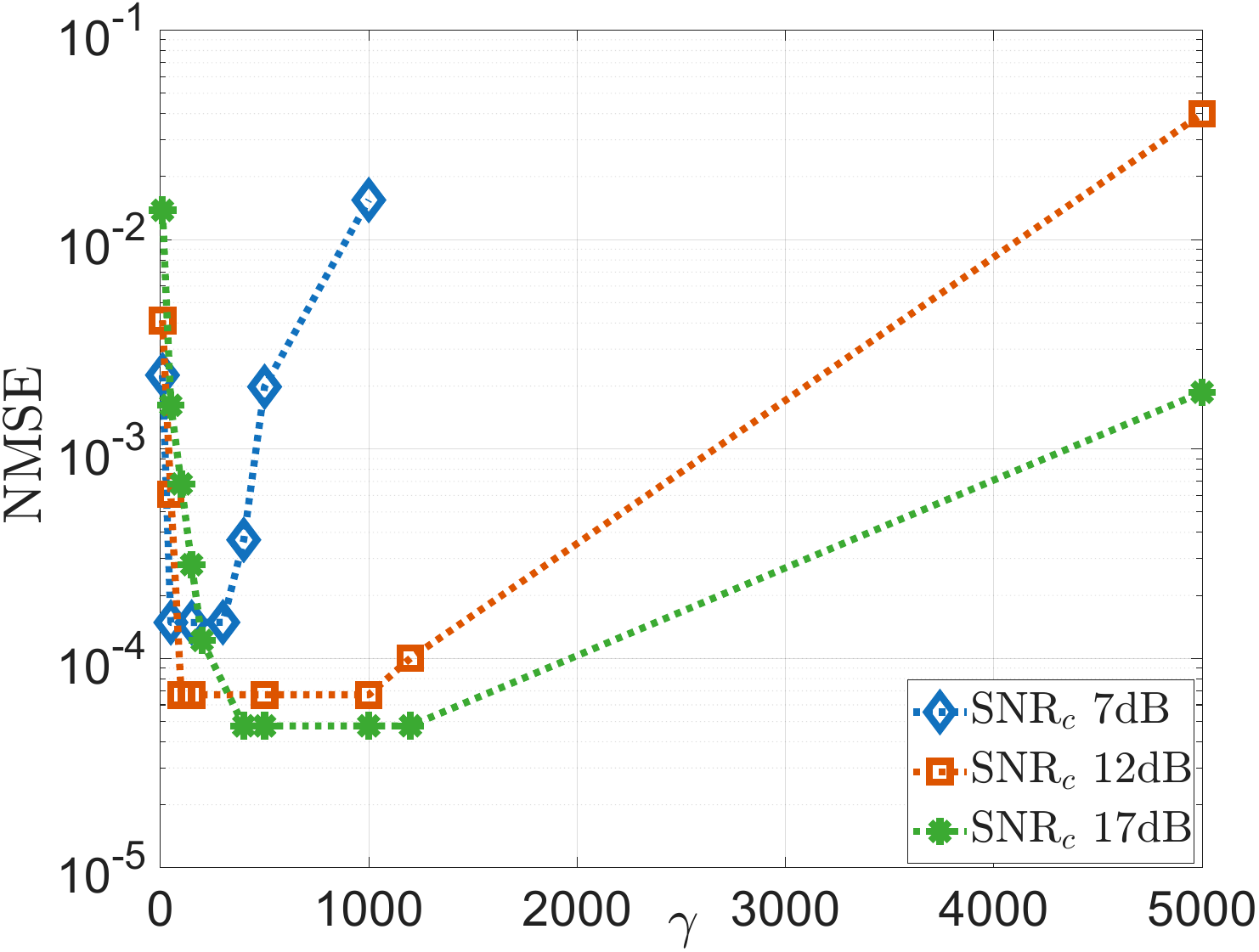}
	 	\caption{\small NMSE vs second stage threshold parameter $\gamma$ for Scenario A}
          \label{nmsegammauniform}
  %\vspace{-.5cm}
	 \end{figure}
  \begin{figure} 
	 	\centering		\includegraphics[width=.9\linewidth]{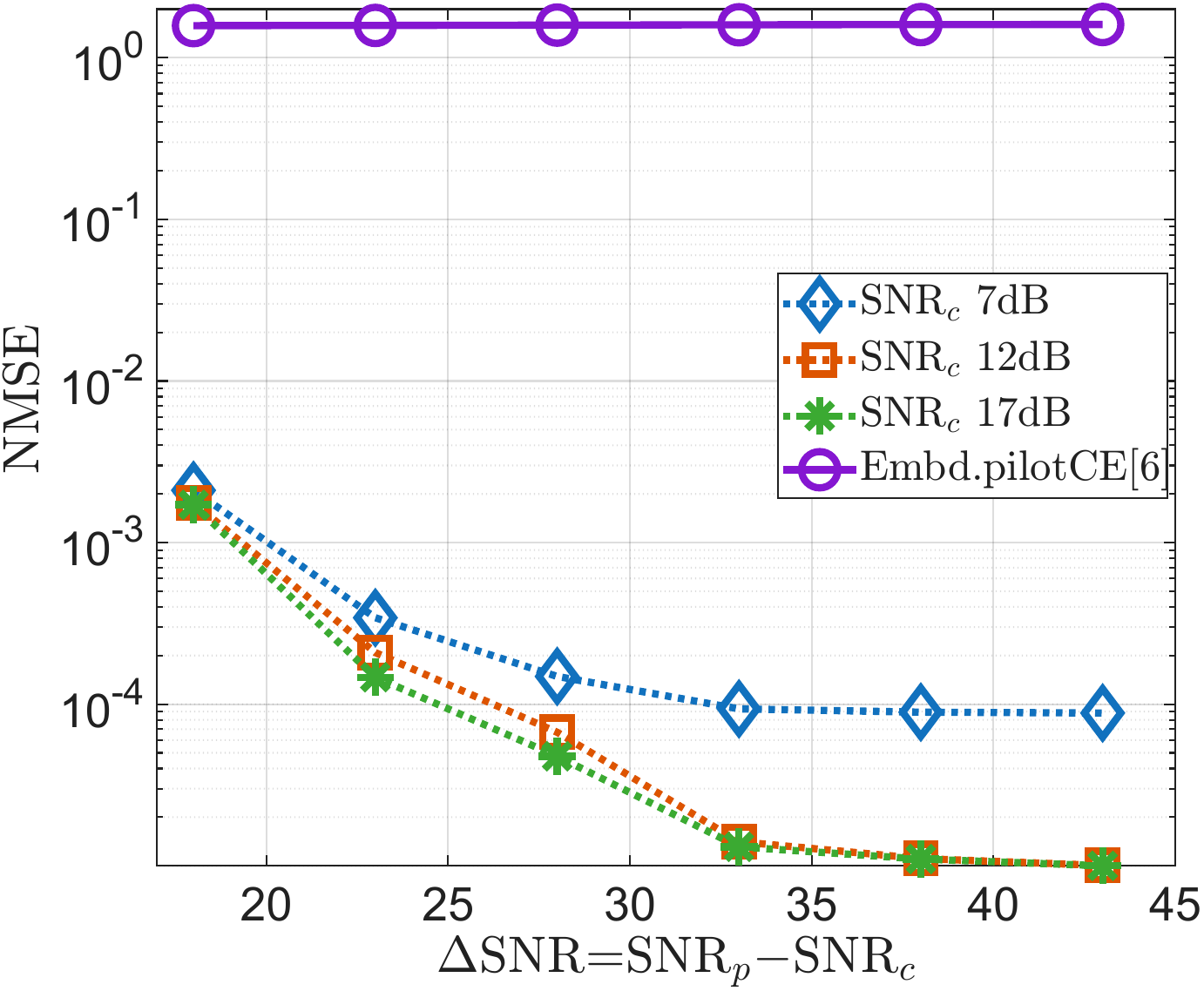}
	 	\caption{\small NMSE vs $\Delta$SNR for Scenario A }%Uniform Channel}
        \label{nmsedeltauniform}
  %\vspace{-.5cm}
	 \end{figure}

       \begin{figure} 
	 	\centering		\includegraphics[width=.9\linewidth]{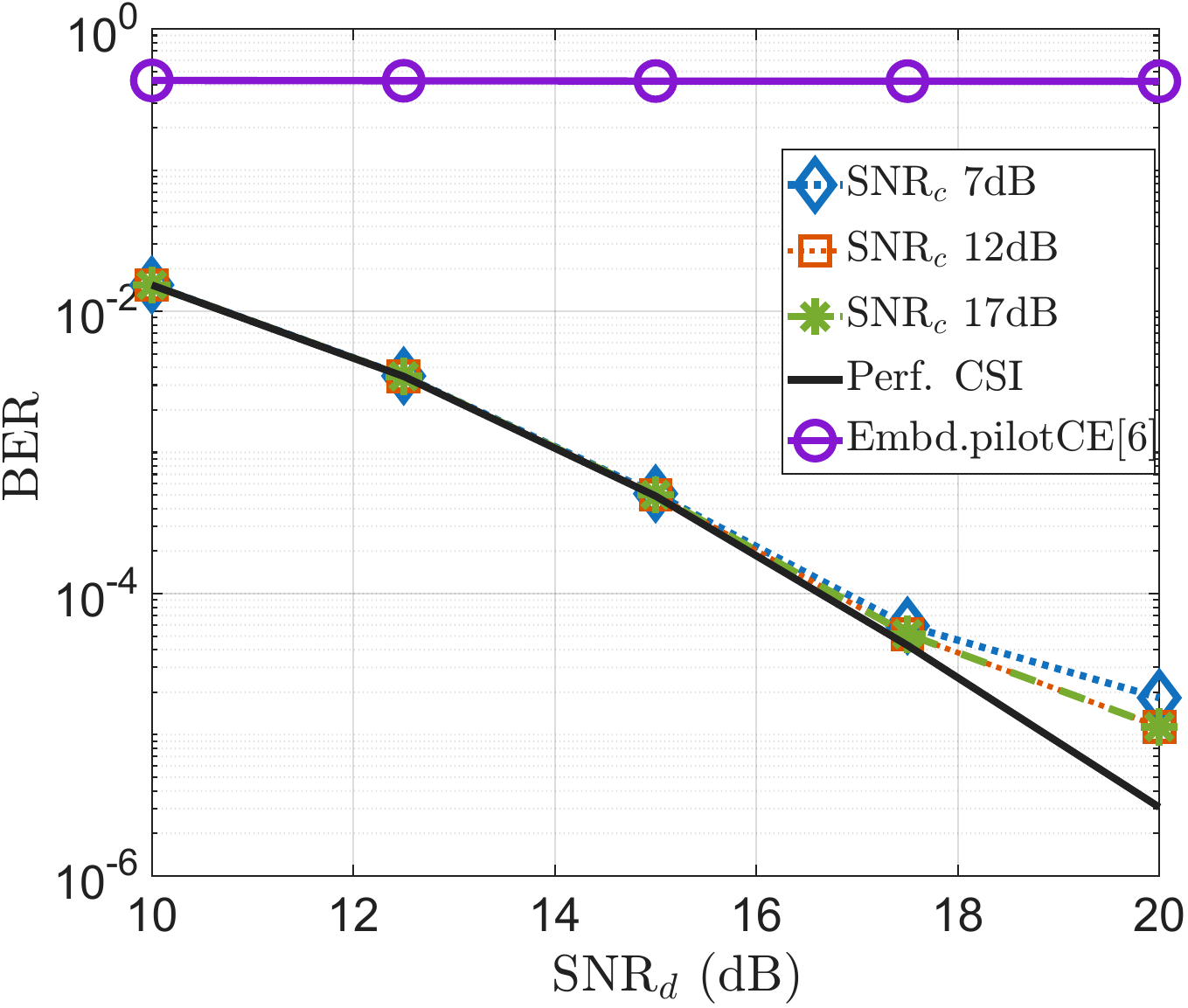}
	 	\caption{\small BER vs $\text{SNR}_{\rm d}$ for Scenario A}
        \label{berdeltauniform}
  %\vspace{-.5cm}
	 \end{figure}
       \begin{figure} 
	 	\centering		\includegraphics[width=.9\linewidth]{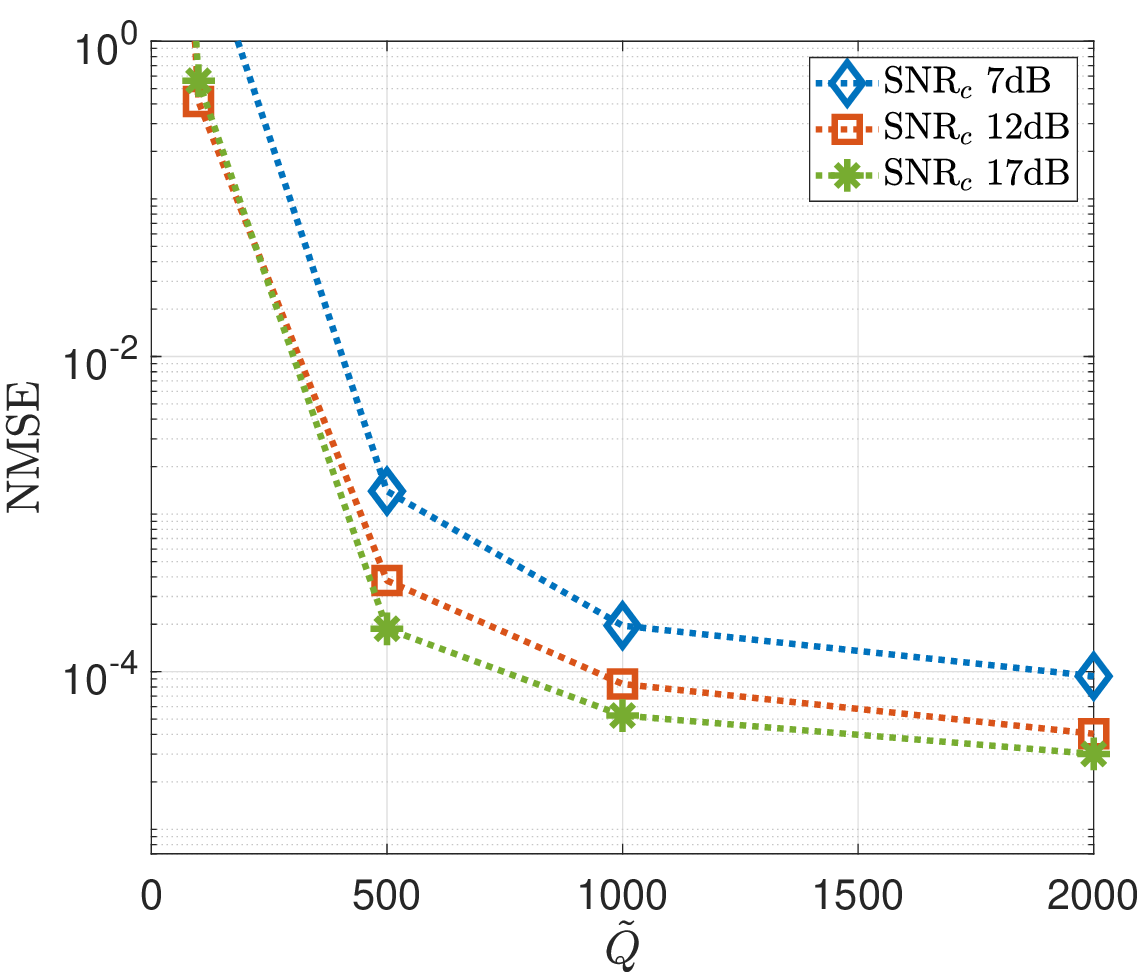}
	 	\caption{\small NMSE vs $\tilde Q$ for Scenario A}
       \label{nmseQtildeuniform}
  %\vspace{-.5cm}
	 \end{figure}

       \begin{figure} 
	 	\centering		\includegraphics[width=.9\linewidth]{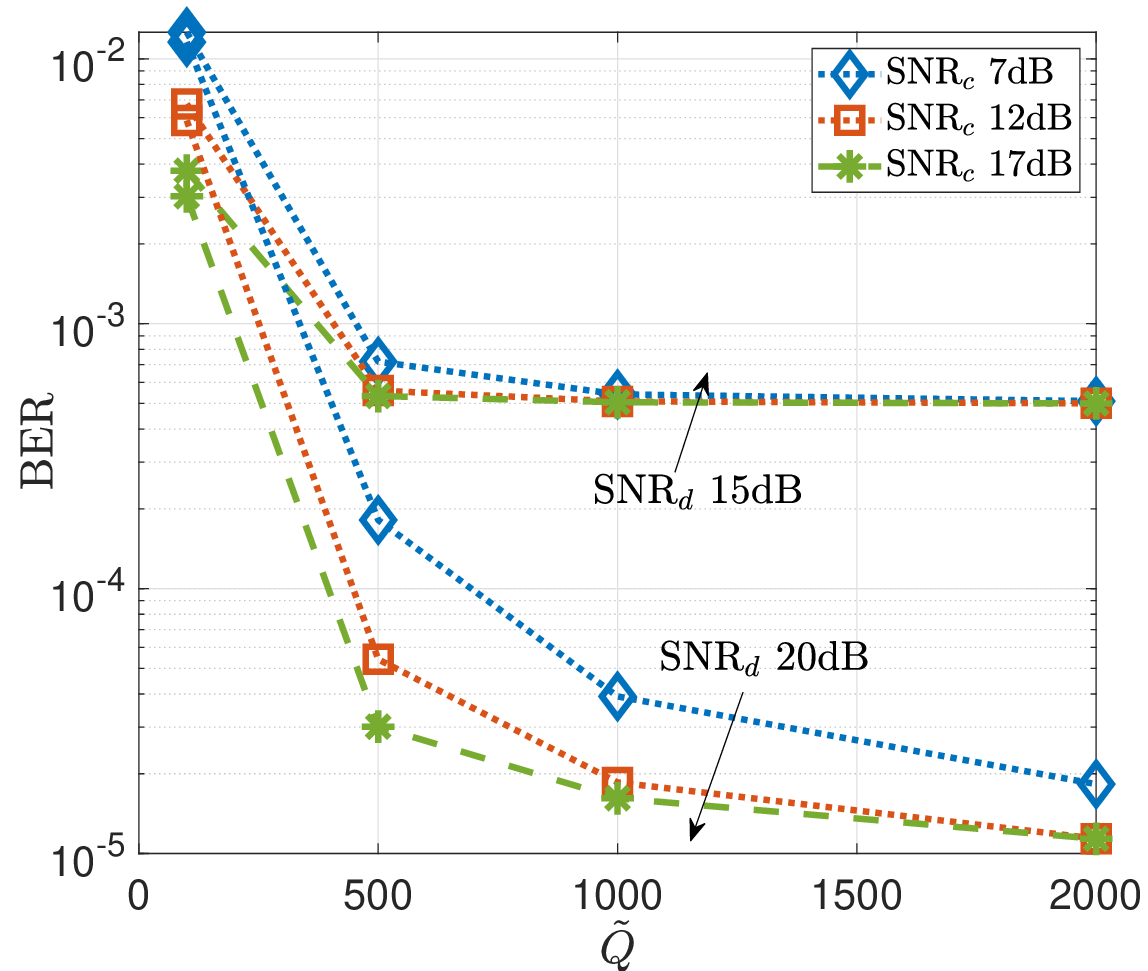}
	 	\caption{\small BER vs $\tilde Q$ for Scenario A}
          \label{berQtildeuniform}
  %\vspace{-.5cm}
	 \end{figure}
%%%%%%%%%%%%%%
Scenario A represents a terrestrial environment with the equal channel gains. It consists of $5$ paths where the delay indices of the paths follows the uniform distribution $\mathcal U(0,20)$ and the Doppler indices follows the uniform distribution $\mathcal U[-160,160]$.

  Fig.~\ref{nmsealphauniform} illustrates the NMSE performance of the proposed channel estimation as a function of the first-stage threshold parameter $\alpha$ for different values of $\text{SNR}_c$ under Scenario~A. The $\Delta$SNR is fixed to $28$~dB. When $\alpha$ is small, the detection threshold is low, which leads to spurious path detections caused by noise and the cosine signal in the DD domain. On the other hand, large values of $\alpha$ result in missed detection of weak paths, which also degrades the NMSE performance. Hence, an intermediate value of $\alpha$ provides the best performance by balancing these two effects. For the considered system parameters, the minimum NMSE is observed around $\alpha = 2.5$ across all $\text{SNR}_c$ values. We further observe that this choice of $\alpha$ remains effective for other $\Delta$SNR values as well considered in this work (results not shown for brevity). Accordingly, $\alpha = 2.5$ is used in all simulations for Scenario~A.

We next present the NMSE performance of the proposed channel estimation as a function of the second-stage threshold parameter $\gamma$ for different values of $\text{SNR}_c$ in Fig.~\ref{nmsegammauniform} for Scenario A. Fix $\Delta$SNR$=28$~dB. Small values of $\gamma$ lead to the detection of noise-dominated frequency components in addition to the actual Doppler peaks, which increases the number of candidate Doppler shifts and degrades the Doppler pairing process, resulting in higher NMSE. On the other hand, for large values of $\gamma$, the detection threshold increases and some low-power Doppler components may not be detected. Hence, an intermediate value of $\gamma$ provides the best NMSE performance. For the considered system parameters, the minimum NMSE is achieved around $\gamma = 300$ across all $\mathrm{SNR}_c$ values. Accordingly, $\gamma = 300$ is used in all subsequent simulations for Scenario~A. We note that the choice of $\gamma$ is independent of $\Delta$SNR, as the second-stage frequency-domain peak detection is performed after removing the single pilot component.

In Fig. \ref{nmsedeltauniform}, we compare the NMSE performance of the proposed two-stage channel estimation as a function of $\Delta\text{SNR}$ under Scenario A for $\text{SNR}_{\rm c} =7$, $12$, and $17$dB, respectively. For all cases, the NMSE reduces with increasing $\Delta$SNR upto $33$dB, thanks to  the improved accuracy on path detection. %However, beyond $\Delta$SNR of 33dB,  it saturates. %, {\color{red} as a further increase in the pilot $\text{SNR}_{\rm p}$ no longer improves path detection due to the presence of noise.} 

The NMSE of $10^{-4}$ provides a sufficiently accurate channel estimation to achieve reliable data detection. This can be achieved at $\Delta$SNR$=33$, $28,~25$dB for $\text{SNR}_{\rm c} =7$, $12$, and $17$dB, respectively. Furthermore, the NMSE improves with increasing $\text{SNR}_{\rm c}$ for $\Delta$SNR $\ge 18$dB, since a higher $\text{SNR}_{\rm c}$ with a sufficient high $\text{SNR}_{\rm p}$ enhances channel estimation in both stages, particularly the second stage when pairing the estimated actual Doppler with delays/channel gains as well as the refinement of channel gain estimation.   Furthermore, we observe that the embedded pilot channel estimation \cite{embedded_pilot_raviteja}, designed specifically for underspread channels, fails completely when applied to overspread channel estimation.

%We also observe that $\Delta$SNR (or equivalently $\text{SNR}_{\rm p}$) should be sufficiently high 

In Fig. \ref{berdeltauniform}, we illustrate the BER performance of ZP-OTFS \cite{tharaj_rake_MRC_journal} using the MRC detection as a function of $\text{SNR}_{\rm d}$ for $\text{SNR}_{\rm c} =7$, $12$, and $17$dB, respectively. We fixed the $\Delta$SNR$=28$dB and the number of observations utilized in the second stage channel estimation $\tilde Q$ to 2000. It is shown that, up to BER of $10^{-4}$, our channel estimation method aligns well with the performance of perfect channel state information (CSI). Around the BER of $2\times 10^{-5}$, $\text{SNR}_{\rm c}=12$ and $17$dB yield the same BER performance with $1$dB gap from perfect CSI. This indicates that $\text{SNR}_{\rm c}=12$dB is sufficient.  In addition, the BER performance utilizing embedded pilot based channel estimation \cite{embedded_pilot_raviteja} completely degrades in overspread channel, as expected.

%, thanks to the reduced NMSE. However, at low $\text{SNR}_{\rm c}$, the noise dominates and thus no significant performance improvement. %Additionally, the cases using $\text{SNR}_{\rm c}=12,17$dB perform similarly, around $1$dB gap from the one with perfect CSI at the BER$=10^{-5}$.

In Fig. \ref{nmseQtildeuniform} and Fig. \ref{berQtildeuniform}, we illustrate the NMSE and BER performance, respectively, as a function of $\tilde Q$ for $\text{SNR}_{\rm c} =7$, $12$, and $17$dB. Recall that $\tilde Q$ denotes the number of time-domain observations used to estimate the channel gains (see Step 7 of the second stage channel estimation), corresponding to paths that share the same delay and aliased Doppler shift.
For BERs in Fig.~\ref{berQtildeuniform}, we adopt $\text{SNR}_{\rm d}=15$, and $20$dB. As expected, both NMSE and BER performance improve with increasing $\tilde Q$, and tend to saturate beyond $\tilde Q = 1000$ across all cases. Hence, $\tilde Q = 1000$ is sufficient to achieve reliable performance in terms of both NMSE and BER.

%%%%%%%%%%%%%%
\subsection{Scenario B}
%%%%%%%%%%%%%%
  \begin{figure}
	 	\centering		\includegraphics[width=1\linewidth]{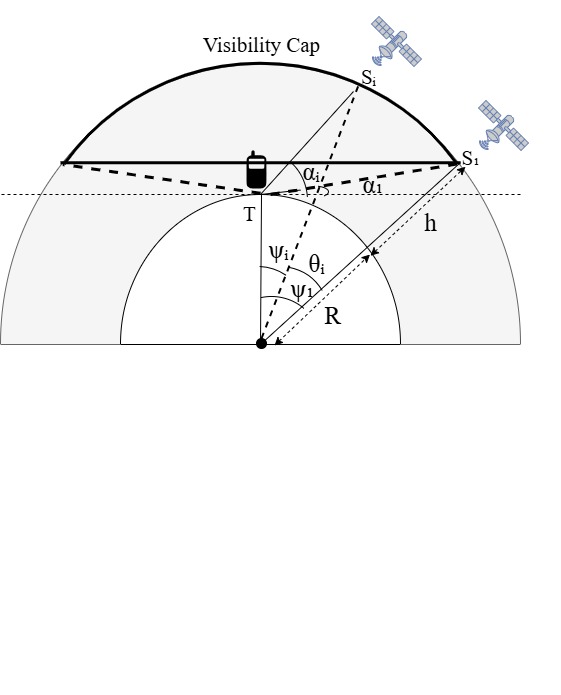}
        \vspace{-5.5cm}
	 	\caption{\small Scenario B}
        \label{leoconst}
 % \vspace{-1cm}
	 \end{figure}
Scenario B considers the downlink multiple LEO Satellites (see Fig. \ref{leoconst}) cooperatively communicating to a ground user (denoted by T). We assume that $S$ satellites are equally spaced along the orbit and fly at the velocity $v_s$ and the altitude $h$ in the user/terminal's {\em visibility cap} (marked by the thicker arc). We represent the earth radius as $R$.

%%%%%%%%%%%%%%%%%%%%%%%
    
Let the satellite $S_1$ (refer Fig. \ref{leoconst}) has an elevation angle $\alpha_1$ with respect to the user/terminal horizon plane, with a  satellite central angle $\psi_1$. Then the angle between satellite $S_1$ to any arbitrary satellite $S_i$, $i\in [1,S]$, is given by 
\begin{equation}
    \theta_i=\frac{2 \pi}{S(R+h)}(i-1)
\end{equation}
The satellite central angle between the $i$-th satellite and the user/terminal can be rewritten as %in terms of $\theta_i$ and satellite central angle of $S_1$ as
\begin{equation}
    \psi_i=\begin{cases}\psi_1-\theta_i, & \theta_i<\psi_1 \\\theta_i-\psi_1, & {\text{otherwise}}\end{cases}
\end{equation}
where $i\in [1,S]$, which can be computed as \cite{ICC_Leo_MIMOOTFS} %{\color{red} should cite Amit ICC paper about LEO sat}
    \begin{equation}
        \psi_i=\text {cos}^{-1} \Big(\frac{R~ \text {cos}\alpha_i}{R+h}\Big)-\alpha_i.
    \end{equation}
%%%%%%%%%%%%%%%%%%%%%%%%%
       \begin{figure} 
	 	\centering		\includegraphics[width=.9\linewidth]{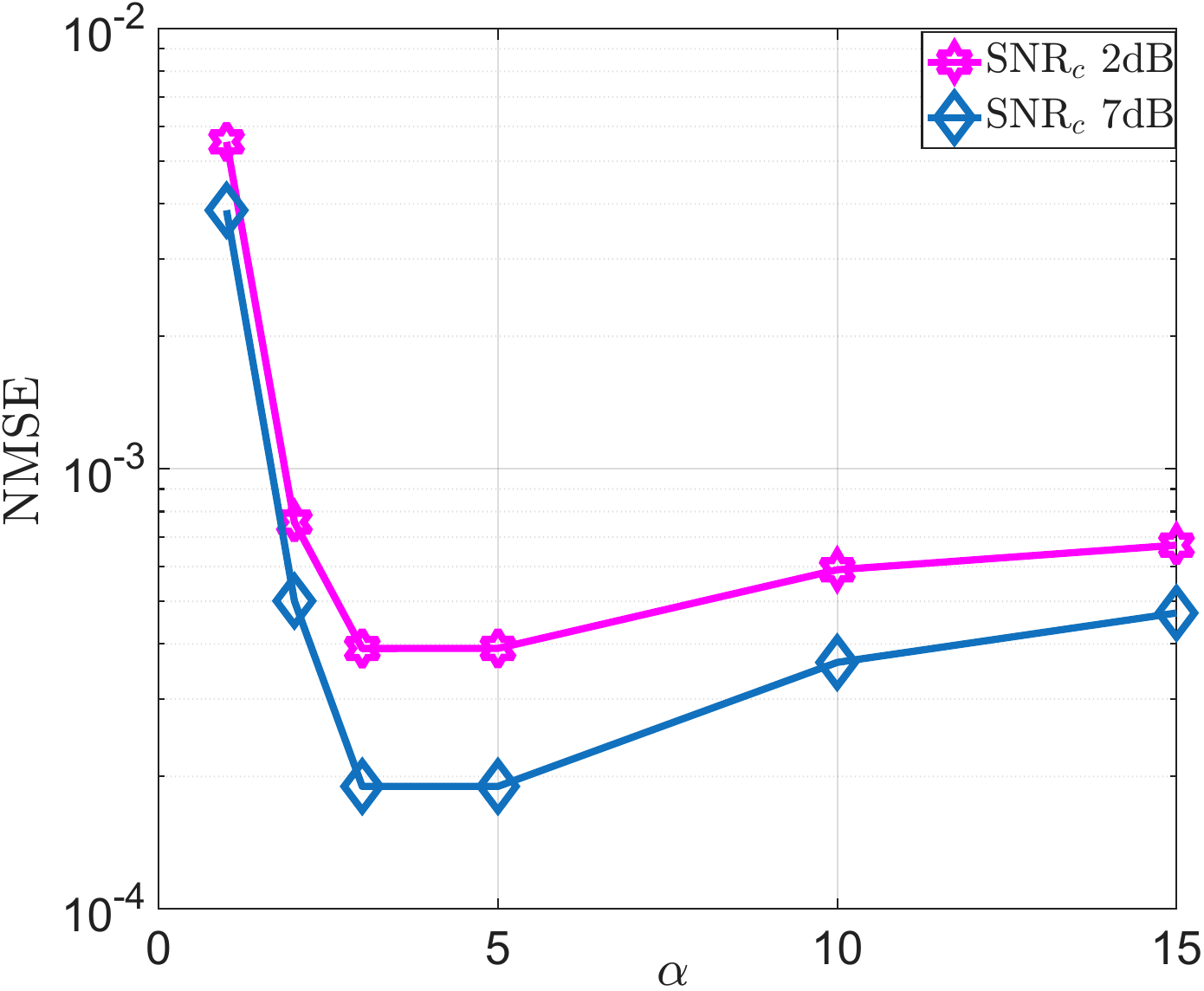}
	 	\caption{\small NMSE vs first stage threshold parameter $\alpha$ for Scenario B}
          \label{nmsvsealphaleo}
  %\vspace{-.5cm}
	 \end{figure}
       \begin{figure} 
	 	\centering		\includegraphics[width=.9\linewidth]{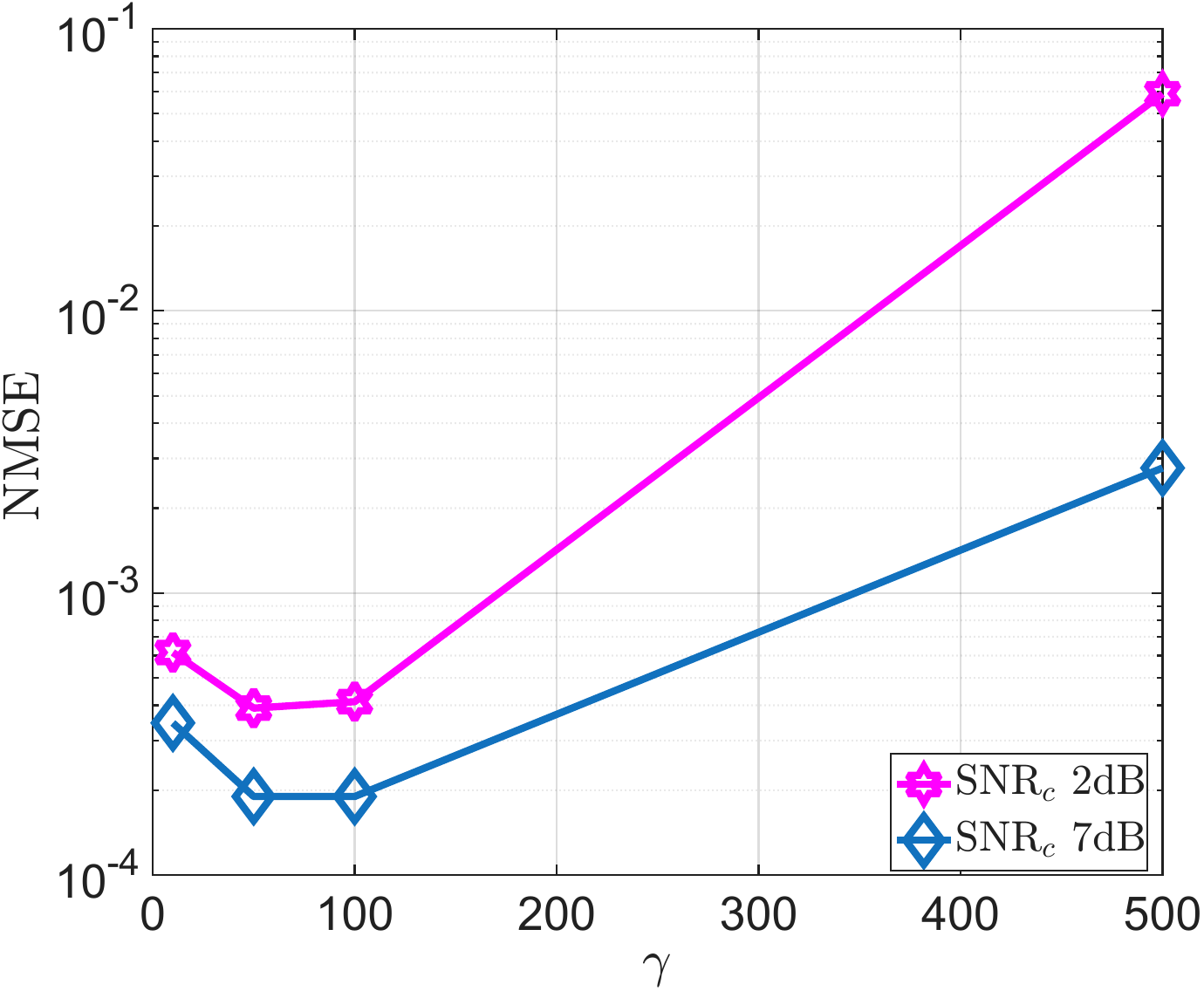}
	 	\caption{\small NMSE vs second stage threshold parameter $\gamma$ for Scenario B}
          \label{nmsevsgammaleo}
  %\vspace{-.5cm}
	 \end{figure}
       \begin{figure} 
	 	\centering		\includegraphics[width=.9\linewidth]{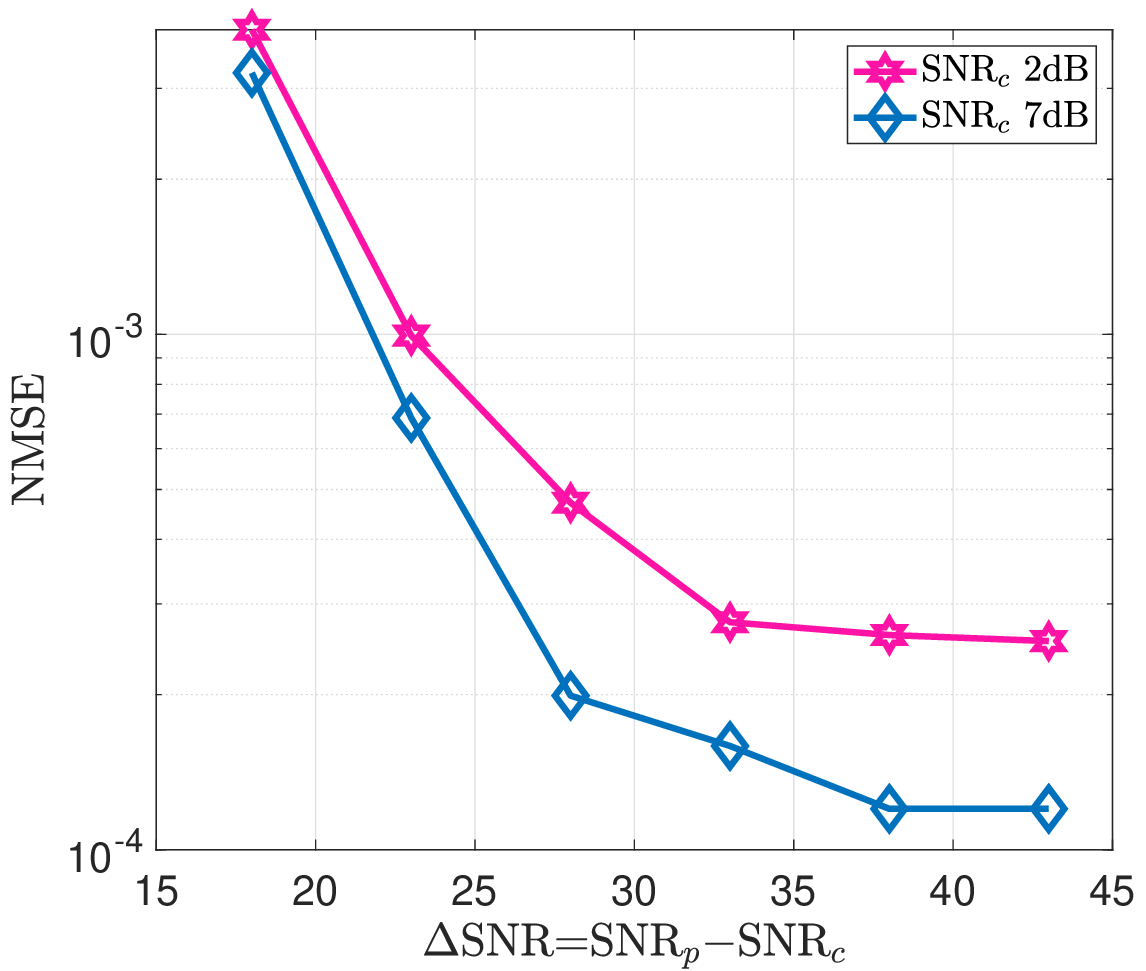}
	 	\caption{\small NMSE vs $\Delta$SNR for Scenario B}
         \label{nmsedeltaleo}
  %\vspace{-.5cm}
	 \end{figure}
%%%%%%%%%%%%%%%%%%%%%%%%%%%%%%%%%
%%%%%%%%%%%%%%%%%%%%%%%%%
       \begin{figure} 
	 	\centering		\includegraphics[width=.9\linewidth]{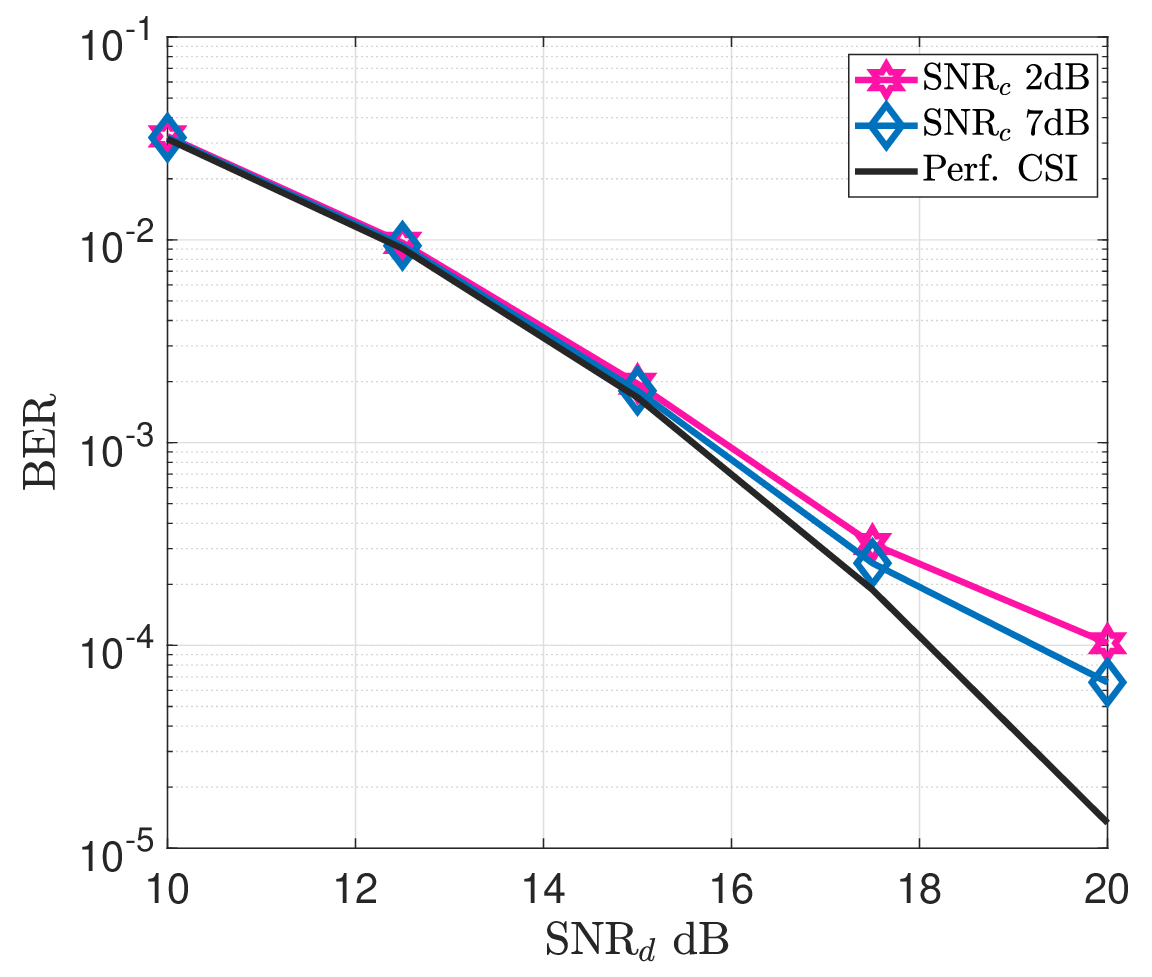}
	 	\caption{\small BER vs data SNR for Scenario B}
         \label{bersnrdleo}
  %\vspace{-.5cm}
	 \end{figure}
%%%%%%%%%%%%%%%%%%%%%%%%%%%%
     
In the user's visibility cap, we assume $L$ number of satellites cooperatively serve the user simultaneously, where $L\le S$.  
Considering a LoS link from each satellite to user, the channel can be modeled in the DD domain as 
\begin{equation}
    h(\tau,\nu)=\sum_{p=1}^{L} h_p \delta(\tau-\tau_p)\delta(\nu-\nu_p)
\end{equation}
where $h_p, \tau_p, $ and $\nu_p$ are the channel gain, propagation delay and Doppler shift associated with the $p$-th satellite, where $p\in[1,L]$.  The delay, Doppler shift, and channel gain can be modeled as follows:

\textbf{Delay:} The propagation delay between satellite $S_p$ and the user is given by
\vspace{-.05cm}
\begin{equation}
    \tau_p=\frac{{d_p}}{c}
\end{equation}
where $d_p={\rm{TS_p}}=\sqrt{R^2+(R+h)^2+2R(R+h)\text{cos}\psi_p}$ is the distance between the $p$-th satellite and the terminal T.

\textbf{Channel Gain:} The channel gains of the paths are modeled as Rayleigh distributed with power of an $p$-th path as
\begin{equation}
    E_p \propto d_p^{-2}
\end{equation}

\textbf{Doppler shift:} The Doppler shift $\nu_p$ of the $p$-th path is computed as
\begin{equation}
    \nu_p=f_c \frac{v_s}{c}\frac{R}{R+h}\cos(\alpha_p) 
\end{equation}

For simulation, we consider $S=120, \alpha_{\rm min}=10^{\circ}, \alpha_{\rm max}=90^{\circ}$. Further, we have $h=550 \text{km}, v_s=7.8$ km/s, and $ R=6378$ km. 
%%%%%
 \begin{figure} [H]
	 	\centering		\includegraphics[width=.9\linewidth]{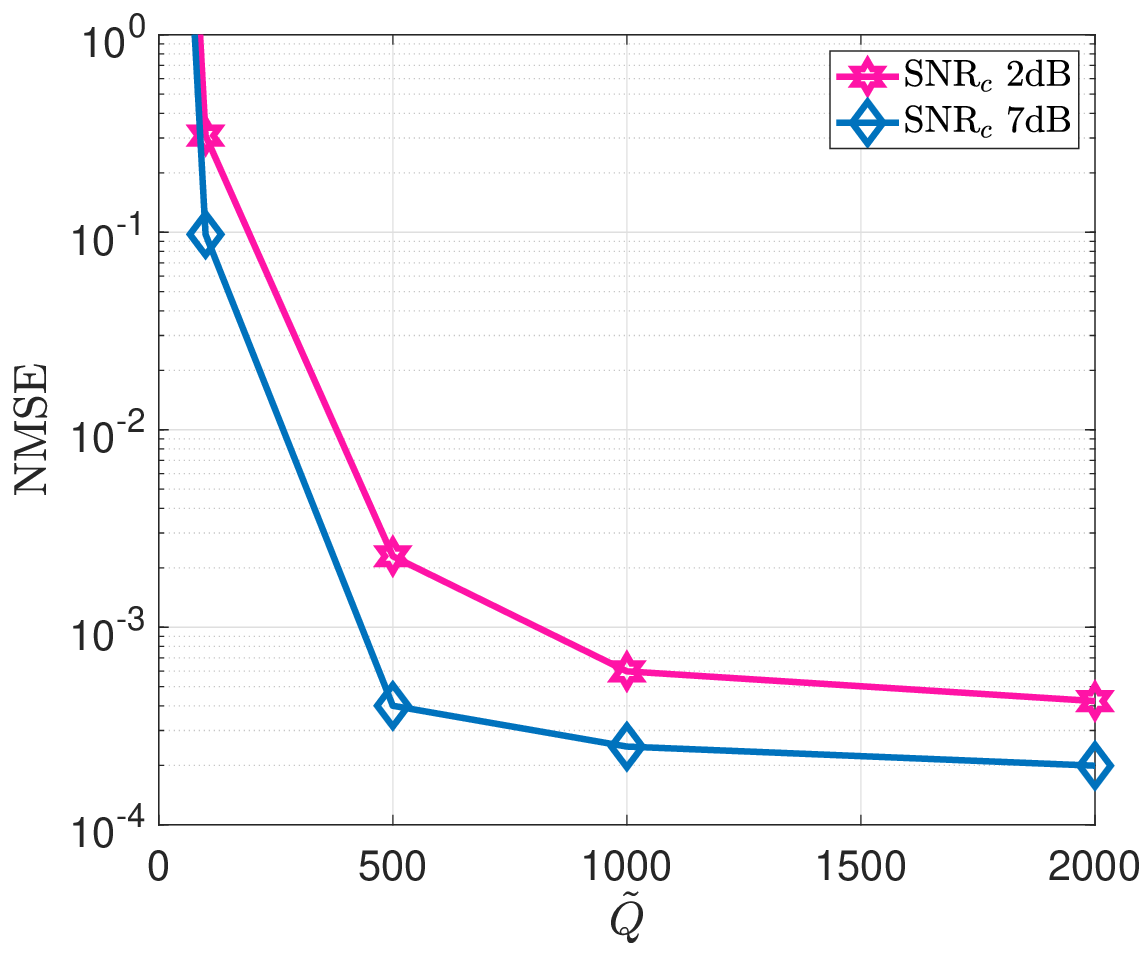}
	 	\caption{\small NMSE vs $\tilde Q$ for Scenario B}
      \label{nmseQtildeleo}
  %\vspace{-.5cm}
	 \end{figure}
       \begin{figure} [H]
	 	\centering		\includegraphics[width=.9\linewidth]{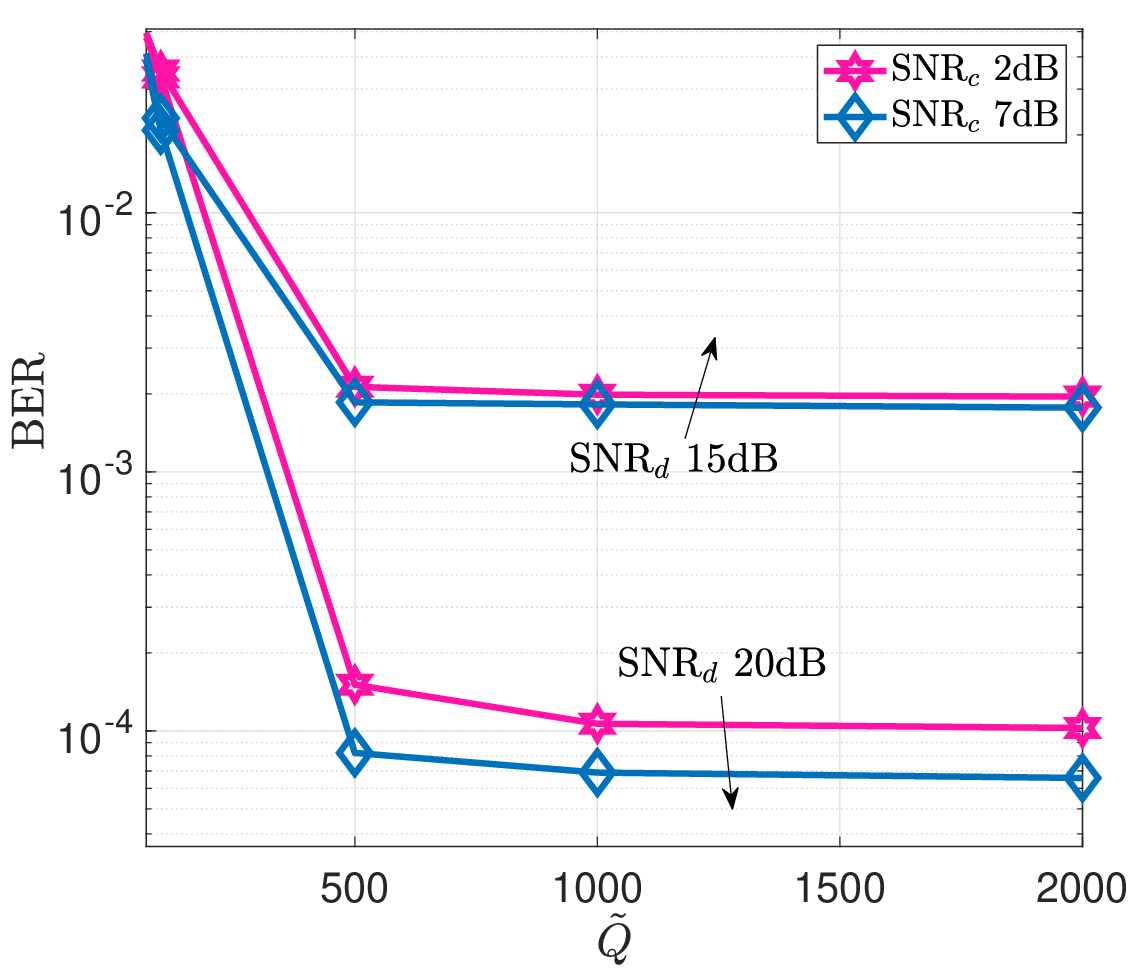}
	 	\caption{\small BER vs $\tilde Q$ for Scenario B}
         \label{berQtildeleo}
  %\vspace{-.5cm}
	 \end{figure}     %%%%%%%%%%%%%%%%%%%%%%%%%%%%%
  We first study the impact of the threshold parameters $\alpha$ and 
$\gamma$ on the NMSE performance for $\text{SNR}_c$ of 2~dB and 7~dB at $\Delta$SNR of 28 dB in Figs.~ \ref{nmsvsealphaleo} and \ref{nmsevsgammaleo}, respectively. The observed trends are consistent with those in Scenario~A. In particular, the minimum NMSE is achieved around $\alpha=3$ and $\gamma=50$. Hence, these parameter values are used in all subsequent simulations for Scenario~B.   

Fig. \ref{nmsedeltaleo} shows the NMSE performance vs $\Delta$SNR of the proposed two-stage channel estimation for Scenario B at $\text{SNR}_{\rm c}$ of $2$ and $7$dB. We obtain similar observations to that of Fig. \ref{nmsedeltauniform}.%, i.e., NMSE reduces with increasing  $\text{SNR}_c$ thanks to the improved path identification and NMSE saturates beyond $\Delta$SNR$=33$dB. We require pilot SNR to be higher than cosine pilot signal SNR by 28dB to attain NMSE greater than $5\times 10^{-4}$.

In Fig. \ref{bersnrdleo}, we illustrate the BER vs $\text{SNR}_{\rm d}$ at $\text{SNR}_{\rm c} =2,~7$dB for Scenario B. We adopt $\Delta$SNR=28dB and $\tilde{Q}=2000$. We obtain similar observations to that of Fig. \ref{berdeltauniform}, where the BER improves with increasing $\text{SNR}_{\rm c}$ and  BER of $10^{-4}$ can be achieved at the $\text{SNR}_{\rm d}=20$dB for $\text{SNR}_{\rm c}=2$dB. Furthermore, compared to the perfect channel, the performance is degraded by $1.5$dB when compared to the one using $\text{SNR}_{\rm c}=7$ dB at BER$=10^{-4}$. 

In Fig. \ref{nmseQtildeleo} and Fig. \ref{berQtildeleo}, we show the NMSE and BER performance, respectively, as function of $\tilde Q$ for $\text{SNR}_{\rm d}$ at $\text{SNR}_{\rm c} =2, 7$dB for Scenario B. We set $\Delta$SNR$=28$dB for all the cases and $\text{SNR}_{\rm d}=15,~20$dB. Similar to Scenario A, both NMSE and BER reduce with increasing $\tilde Q$ and saturates beyond $\tilde{Q}=1000$.

\subsection{ 
Remark: Extension to the Fractional Doppler Scenarios} 

 To incorporate the fractional Doppler estimation, the proposed framework can be extended by employing a high-resolution FFT operation in the second stage of the estimation process. In particular, the first stage estimates the integer part of the aliased fractional Doppler values using the proposed methodology, while the second stage estimates the actual fractional Doppler value through a high-resolution spectral analysis. Thereafter, the same pairing methodology used for the integer delay-Doppler case can be employed to associate the corresponding delay and fractional Doppler pairs. For fractional Doppler pairing, the set in (\ref{eq:aliasedDoppset}) is redefined as $\mathbb K=\left\{\mathbbm k_{i}=\left\lfloor[\hat k_{i}]_N\right\rfloor , \text{for~all}~\hat k_{i}\in \mathcal K\right\}$ since the estimate in the first stage is an integer value.

   Furthermore, due to the leakage introduced by fractional Doppler components, the channel coefficients estimated directly in the delay-Doppler domain become inaccurate. Hence, the channel coefficients can be re-estimated in the time domain using (\ref{chann_Est_LS}).

 The above extension works well for Case 3 discussed in Section III-B, where the paths have distinct delays and distinct (apart by atleast 1 grid) aliased fractional Doppler values. However, for Case 2, i.e., distinct delays but close aliased fractional Doppler values, the estimation becomes inaccurate for approximately $1\%$ of the channel realizations. This is because the proposed pairing methodology for associating the delay with the actual Doppler value in Case 2 relies on the channel coefficient estimates in the delay-Doppler domain, which become inaccurate in the presence of fractional Doppler. Consequently, the pairing performance degrades. Moreover, the estimation becomes highly inaccurate for Case 1, i.e.,
paths with identical delay and aliased Doppler values due to the involved complexity and inaccurate
delay-Doppler channel coefficient estimate.

Resolving Cases 1 and 2 requires further investigation, which is currently beyond the scope of this paper. These limitations will be addressed in our future work.

\section{Conclusion}\label{Sec:Concl}
In this paper, we investigate an OTFS system in time-varying channels with overspread Doppler shifts. The overspread Doppler shifts can result in aliased Doppler shifts in the DD domain due
to the OTFS modulo $N$ operation, which hampers standard channel estimation methods. To address this problem, 
we propose a DD training frame and a two-stage channel
estimation method, where the training frame consists of a cosine pilot signal and a pilot symbol. Based on this, we develop a two-stage channel estimation algorithm. In the first stage of the channel estimation, we use the
pilot symbol in the DD domain to estimate the delays, aliased Doppler shifts, and channel gains of the propagation
paths. In the second stage, the received time domain signal is converted into the frequency domain for peak detection, identifying all the Doppler shifts of the cosine pilot signal frequency. Then, we present a threshold-based method to pair the estimated actual Doppler shifts with their corresponding delays and channel gains. We further analyze the computational complexity of the proposed channel estimation method.  
Finally, we demonstrate the effectiveness of our channel estimation in terms of NMSE and BER performance for different scenarios.   Our future work will extend the proposed algorithm to account for fractional delay and Doppler effects. 
%%%%%%%%%%%%%%%%%%%%%%%%%
%%%%%%%%%%%%%%%%%%%%%%%%
%%%%%%%%%%%%%%%%%%%%%%%%%%%%%%%%%%
\section*{Appendix}

% \subsection{Proof of Lemma 1}
% The normalized Doppler shift for a channel path is modeled as a random variable 
% \begin{equation}
%     \kappa = k_{\max} \cos\theta,
% \end{equation} 
% which is a function of a random variable 
% $\theta \sim \mathcal{U}[0,2\pi)$.  The probability density function of the Doppler shift is  
% \begin{equation}
%     f_{\kappa}(x) = \frac{1}{\pi k_{\max} \sqrt{1 - \left( \frac{x}{k_{\max}} \right)^2}}, \quad -k_{\max} \leq x \leq k_{\max}.
% \end{equation}  

% A path will have a normalized {\em integer} Doppler shift $k = \lfloor k_{\max} \cos \theta \rceil$, that is, when
% \begin{equation}
%     k - 0.5 \leq k_{\max} \cos \theta < k + 0.5.
% \end{equation}  
% Thus, the probability that a path has an integer Doppler shift \( k \) is given by  
% \begin{equation}
%     p_k = \int_{k-0.5}^{k+0.5} f_{\kappa} (x) \, dx.
% \end{equation}  

% The probability that exactly \( p \) out of \( L \) paths attain an integer Doppler shift \( k\) where \(k\in [-k_{\max}, k_{\max}] \) is  
% \begin{equation}\label{eq:qp}
%     q_p = \sum_{k=-k_{\max}}^{k_{\max}} \binom{L}{p} p_k^p (1 - p_k)^{L - p}.
% \end{equation}  

% Thus, the probability that no two paths have the same integer Doppler shift is given by
% \begin{align}
%     P_r &= 1-{\text Pr}(\text{2 paths have the same Doppler}) \\
%         &~~~~~- {\text Pr}(\text{3 paths have the same Doppler}) \\
%         &~~~~~~~~~~~~~~~~~~~~~~~~~\vdots \\
%         &~~~~~- {\text Pr}(L \text{ paths have the same Doppler}) \\
%         &= 1 - \sum_{p=2}^{L} q_p.
% \end{align}  

%  \vspace{0.2 cm}

 \subsection{Proposed Second Stage CE Algorithm}\label{Appen:Alg}
  \hrule
 \vspace{0.05 cm}
 {\bf{ Stage 2 CE Based on Cosine Peak in Frequency Domain}} 
% \vspace{0.05 cm}
 \hrule
% \vspace{0.1 cm}
 \begin{algorithmic} [1]
    			%\STATE {Obtain
        % $\hat l_{\hat {\ell}_ij}=\hat {\ell}_i+b_\betaM, ~~\forall b_j\in \mathcal B$}
%        \STATE {\mathbf{Input:} $\mathbf r_{\rm t}, \mathbf s_{\rm t}, \gamma, \sigma_w^2,\mathcal B$.}
    %    \STATE {$\mathbf{Input:} :~{\text{from first step CE}}: {\mathcal J},  \mathcal K_{\hat \ell_\mu} ~\forall \mu, \mathcal H$}.
    %   \STATE{\mathbf{Output:} ${\mathcal H} =\{(\hat l_j, \hat k_j, \hat h_j)\}$}
    \STATE{$\text{\bf func}~[{\mathcal H}]=\textit{CosineFreqPeak} (\mathbf r_{\rm t}, \gamma,  \delta, x_p, \sigma_w^2, f_0,M,  N,$ \\
    ~~~~~~~~~~~~~~~~~~~~~~~~~~~~~~~~~~~~~~~~~~~~~~~~~~~~~~~~~~~~$ \mathcal H,\mathcal A,\mathcal L_{\hat{k}'_{\iota}},  \mathcal G_{\hat{k}'_{\iota}}, \forall \iota$)}
    %   \STATE {Compute $\mathcal B$ using (Steps a-c).}
     \STATE {Set $\mathcal K=\{\},  i=1, j=1$}
%      \FOR{$q\rightarrow0~\text{to}~MN-1$}
 %     \STATE{$\check{\mathbf r}_{\rm c}[q]\leftarrow \mathbf r_{\rm t}[q]$}
%      \IF{$|\mathbf r_{\rm t}[q]|^2>\Gamma=\gamma_1 |x_p|^2$}
%\STATE{$\check{\mathbf r}_{\rm c}[q]\leftarrow 0$}
%       \ENDIF
 %      \ENDFOR
 \STATE{Obtain $\check{\mathbf r}_{\rm c}$ using (\ref{rxDtcosine}) and (\ref{rxcheckcosine}).}
       \STATE{Obtain $\mathbf R=\mathbf F_{MN}\check{\mathbf r}_{\rm c}$}
      % \For {$f\rightarrow 0~\text{to}~\frac{MN}{2}$}
       
      % \ENDFOR
       
       \FOR{$f\rightarrow 0~\text{to}~\frac{MN}{2}$}
      \IF{$ |\mathbf R[f]|^2>  \gamma~\sigma_w^2$}
       \STATE{ $\hat k_{i}=f-f_0N, \mathcal K=\mathcal K\cup \{\hat k_{i}\}$}
       \STATE{$i \leftarrow i+1$}
       \ENDIF
       \ENDFOR
       \STATE{$\mathbb K=\{\mathbbm k_{i}=[\hat k_{i}]_N\},\hat k_{i}\in \mathcal K $}
    \STATE{$\tilde {\mathbb K}=Unique(\mathbb K)$}
       \FOR{$\tilde i\rightarrow 1~\text{to}~|\tilde{\mathbb K}|$}
       \STATE{$\beta=0$}
       \FOR{$i \rightarrow 1~\text{to}~|\mathbb K|$}
       \IF{$\mathbbm k_{\tilde i}\in \tilde{\mathbb K} =\mathbbm k_{i}\in {\mathbb K}$}
        \STATE{$ \beta=\beta+1, ~ \mathbb K_{{\tilde i}}=\{\}$}
        \STATE{$\hat k_{\beta}(\mathbbm k_{\tilde i})=\hat k_{i}, \mathbb K_{\tilde i}=\mathbb K_{{\tilde i}}\cup  \hat k_{\beta}(\mathbbm k_{\tilde i})$}
        \ENDIF
      \ENDFOR
      \STATE{Find $\iota$ such that $\hat{k}'_{\iota}\in \mathcal A=\mathbbm k_{\tilde i}$}
      \IF{$|\mathbb K_{\tilde i}|=1$}
      \STATE{$\hat l_{j}=\hat l_{1}(\hat{k}'_\iota=\mathbbm k_{\tilde i}), \hat k_{j}=\hat k_{1}(\mathbbm k_{\tilde i}),\hat h_{j}=\hat h_{1}(\hat{k}'_\iota)$}
     % \STATE{Pair $\{\hat l_{1}(\hat{k}'_\iota=\mathbbm k_{\tilde i}), k_{\beta}(\mathbbm k_{\tilde i}), \hat h_{1}(\hat{k}'_\iota)\}$}
       %\STATE{$\mathcal H\leftarrow\mathcal H \cup \{\hat l_{1}(\hat{k}'_\iota=\mathbbm k_{\tilde i}), k_{\beta}(\mathbbm k_{\tilde i}), \hat h_{1}(\hat{k}'_\iota)\}$.}
       \STATE{$\mathcal H\leftarrow\mathcal H \cup \{\hat l_{j}, \hat k_{j}, \hat h_{j}\}$.}
       \STATE{$j=j+1$}
       \ELSIF{$|\mathbb K_{\tilde i}|=|\mathcal L_{\hat{k}'_{\iota}}|>1$}
      % \For {$\lambda \rightarrow 1~\text{to}~|\mathcal L_{\hat{k}'_{\iota}}|$}
%      \FOR{$\beta \rightarrow 1~\text{to}~|\mathbb K_{\tilde i}|$}
       \FOR{$\lambda \rightarrow 1~\text{to}~|\mathcal L_{\hat{k}'_{\iota}}|$}
      \STATE{$\beta^\star(\lambda)=\argmin\limits_{\beta} \big|\big|\frac{2}{A\sqrt{M}}\mathbf R({\hat k}_{\beta}(\mathbbm k_{\tilde i}) +f_0N)\big|- |\hat h_{\lambda}(\hat{k}'_\iota)\big|\big|$}
         \STATE{ $\hat l_{j}=\hat l_{\lambda}(\hat{k}'_\iota=\mathbbm k_{\tilde i}), \hat k_{j} = \hat k_{\beta^*(\lambda)}(\mathbbm k_{\tilde i}), \hat h_{j} = \hat h_{\lambda}(\hat{k}'_\iota)$}
       \STATE{$\mathcal H\leftarrow\mathcal H \cup \{\hat l_{j}, \hat k_{j}, \hat h_{j}\}$.}
         \STATE{$j=j+1$}
    %   \STATE{Pair $\{\hat l_{\lambda}(\hat{k}'_\iota=\mathbbm k_{\tilde i}), k_{\beta}(\mathbbm k_{\tilde i}), \hat h_{\lambda}(\hat{k}'_\iota)\}$}
    %   \STATE{$\mathcal H\leftarrow\mathcal H \cup \{\hat l_{\lambda}(\hat{k}'_\iota=\mathbbm k_{\tilde i}), k_{\beta}(\mathbbm k_{\tilde i}), \hat h_{\lambda}(\hat{k}'_\iota)\}$.}
%       \ENDIF
  %    \ENDFOR
      \ENDFOR
      \ELSIF{$|\mathbb K_{\tilde i}|>1 ~\&~ |\mathcal L_{\hat{k}'_{\iota}}|=1$}
      \FOR{$\beta \rightarrow 1~\text{to}~|\mathbb K_{\tilde i}|$}
       \STATE{ $\hat l_{j}=\hat l_{1}(\hat{k}'_\iota=\mathbbm k_{\tilde i}), \hat k_{j}=\hat k_{\beta}(\mathbbm k_{\tilde i})$}
        \STATE{Set $\hat h_{j}=0$}
         \STATE{$\mathcal H\leftarrow\mathcal H \cup \{\hat l_{j}, \hat k_{j}, \hat h_{j}\}$.}
         \STATE{$j=j+1$}
    %  \STATE{Pair $\{\hat l_{1}(\hat{k}'_\iota=\mathbbm k_{\tilde i}), k_{\beta}(\mathbbm k_{\tilde i})\}$}
%      \STATE{Compute $\hat {\mathbf h}$ using ().}
   %   \STATE{Obtain $\hat h_{\beta}$ corresponding to delay-Doppler pair $\{\hat l_{1}(\hat{k}'_\iota=\mathbbm k_{\tilde i}), k_{\beta}(\mathbbm k_{\tilde i})\}$ from $\hat {\mathbf h}$}
  % \STATE{Initialize $\hat h_{j}=0$
  %     \STATE{$\mathcal H\leftarrow\mathcal H \cup \{\{\hat l_{1}(\hat{k}'_\iota=\mathbbm k_{\tilde i}), k_{\beta}(\mathbbm k_{\tilde i}),\hat h_{\beta}\}$.}
      \ENDFOR
      \ELSIF{$|\mathbb K_{\tilde i}|> |\mathcal L_{\hat{k}'_{\iota}}|>1 $}
%     \STATE{$ \mathcal B=\{1,\ldots, |\mathbb K_{\tilde i}|\}$}
    %  \FOR{$\beta \rightarrow 1~\text{to}~|\mathbb K_{\tilde i}|$}
       \FOR{$\lambda \rightarrow 1~\text{to}~|\mathcal L_{\hat{k}'_{\iota}}|$}
      \STATE{$\beta^\star(\lambda)=\argmin\limits_{\beta} \big|\big|\frac{2}{A\sqrt{M}}\mathbf R({\hat k}_{\beta}(\mathbbm k_{\tilde i}) +f_0N)\big|- |\hat h_{\lambda}(\hat{k}'_\iota)\big|\big|$}
         \STATE{ $\hat l_{j}=\hat l_{\lambda}(\hat{k}'_\iota=\mathbbm k_{\tilde i}), \hat k_{j}=\hat k_{\beta^{\star}(\lambda)}(\mathbbm k_{\tilde i}), \hat h_{j}=\hat h_{\lambda}(\hat{k}'_\iota)$}
           \STATE{$\mathcal H\leftarrow\mathcal H \cup \{\hat l_{j}, \hat k_{j}, \hat h_{j}\}$.}
         \STATE{$j=j+1$}

    %   \STATE{Pair $\{\hat l_{\lambda}(\hat{k}'_\iota=\mathbbm k_{\tilde i}), k_{\beta}(\mathbbm k_{\tilde i}), \hat h_{\lambda}(\hat{k}'_\iota)\}$}
    %   \STATE{$\mathcal H\leftarrow\mathcal H \cup \{\hat l_{\lambda}(\hat{k}'_\iota=\mathbbm k_{\tilde i}), k_{\beta}(\mathbbm k_{\tilde i}), \hat h_{\lambda}(\hat{k}'_\iota)\}$.}
       \STATE{Remove $\hat k_{\beta^{\star}(\lambda)}(\mathbbm k_{\tilde i})$ from $\mathbb K_{\tilde i}$}
       \STATE{Remove $\hat l_{\lambda}(\hat{k}'_\iota=\mathbbm k_{\tilde i})$ from $\mathcal L_{\hat{k}'_{\iota}}$}
        \STATE{Remove $\hat h_{\lambda}(\hat{k}'_\iota=\mathbbm k_{\tilde i})$ from $\mathcal G_{\hat{k}'_{\iota}}$}
   %    \ENDIF
   %    \ENDFOR
       \ENDFOR
        \FOR{$\beta \rightarrow 1~\text{to}~|\mathbb K_{\tilde i}|$}
          \STATE{ $\hat l_{j}=\hat l_{1}(\hat{k}'_\iota=\mathbbm k_{\tilde i}), \hat k_{j}=\hat k_{\beta}(\mathbbm k_{\tilde i})$}
        \STATE{Set $\hat h_{j}=0$}
         \STATE{$\mathcal H\leftarrow\mathcal H \cup \{\hat l_{j}, \hat k_{j}, \hat h_{j}\}$.}
         \STATE{$j=j+1$}
%        \STATE{Pair $\{\hat l_{1}(\hat{k}'_\iota=\mathbbm k_{\tilde i}), k_{\beta}(\mathbbm k_{\tilde i})\}$}
     %   \STATE{$\mathcal H\leftarrow\mathcal H \cup \{\{\hat l_{1}(\hat{k}'_\iota=\mathbbm k_{\tilde i}), k_{\beta}(\mathbbm k_{\tilde i}),\hat h_{\beta}\}$.}
       \ENDFOR
       \ENDIF
       \ENDFOR
        \STATE{Compute $\hat {\mathbf h}$ using (\ref{chann_Est_LS}).}
          \FOR{$j \rightarrow 1~\text{to}~|\mathcal H|$}
          \IF{$\hat h_{j}=0$}
          \STATE{$\hat h_{j}=\hat {\mathbf h}[j]$}
          \ENDIF
          \ENDFOR
   %   \STATE{Obtain $\hat h_{\beta}$ corresponding to delay-Doppler pair $\{\hat l_{1}(\hat{k}'_\iota=\mathbbm k_{\tilde i}), k_{\beta}(\mathbbm k_{\tilde i})\}$ from $\hat {\mathbf h}$}
    
      \STATE{Return $\mathcal H$}
    		\end{algorithmic} \label{Alg_chirp}
   %         \end{algorithm}
      \hrule
       %%%%%%%%%%%%%%%%%%%%%%%%%%%%%%%%%%%%
       \vspace{0.1cm}
%%%%%%%%%%%%%%%%%% 
	\ifCLASSOPTIONcaptionsoff
	\newpage
	\fi
	\bibliographystyle{IEEEtran}
	\bibliography{reference}	

@article{wang2023road6g,
  title="{On the Road to 6G: Visions, Requirements, Key Technologies, and Testbeds}",
  author={Wang, Cheng-Xiang and You, Xiaohu and Han, Tao and Zhang, Hong and Yuan, Da-Cheng and Ai, Bo and Zhang, Jiaheng and others},
  journal={IEEE Communications Surveys $\&$ Tutorials},
  year={2023},
  volume={25},
  number={3},
  pages={1552--1616},
  doi={10.1109/COMST.2023.3257056},
  publisher={IEEE}
}

@INPROCEEDINGS{hadani2017orthogonal,
  author={Hadani, R. and others},
  booktitle={Proc. IEEE Wireless Commun. Net. Conf.}, 
  title="{Orthogonal Time Frequency Space Modulation}", 
  year={2017},
  address={ San Francisco, CA, USA},
  volume={},
  number={},
  pages={1-6},
  doi={10.1109/WCNC.2017.7925924}}

@ARTICLE{EffDiversityLett_Ravi,
  author={Raviteja, P. and Hong, Yi and Viterbo, Emanuele and Biglieri, Ezio},
  journal={IEEE Wireless Commun. Lett.}, 
  title="{Effective Diversity of OTFS Modulation}", 
  year={2020},
  volume={9},
  number={2},
  pages={249-253},
  doi={10.1109/LWC.2019.2951758}}

@INPROCEEDINGS{ICC_Leo_MIMOOTFS,
  author={Bora, Amit Sravan and Phan, Khoa T. and Hong, Yi},
  booktitle={Proc. IEEE Intern. Conf. Commun. Workshops}, 
  title="{Spatially Correlated MIMO-OTFS for LEO Satellite Communication Systems}", 
  year={2022},
  address={Seoul, Korea, Republic of},
  volume={},
  number={},
  pages={723-728},
  doi={10.1109/ICCWorkshops53468.2022.9814666}}

@ARTICLE{VihanOTFSISAC2025_embeddedpilots,
  author={Kalpage, Nisal and Hong, Yi and Priya, Preety },
  journal={IEEE Trans. Veh. Tech.}, 
  title="{Performance Analysis and Joint Receiver Design for an OTFS-Based ISAC System in Time-Varying Channels}", 
  year={2025},
  volume={ },
  number={ },
  pages={1-6},
  doi={10.1109/TVT.2025.3597365}}

@ARTICLE{LuoJiaHe_VLEO,
  author={ Luo, H. and et. al.},
  journal={Communications of Huawei Research}, 
  title="{Very-Low-Earth-Orbit Satellite Networks
for 6G}", 
  year={2022},
  volume={ },
  number={2},
  pages={40-54},
  }

@ARTICLE{Shi_TWC_OTFS,
  author={Shi, D. and Wang, W. and You, L. and Song, X. and Hong, Y. and Gao, X. and Fettweis, G.},
  journal={IEEE Trans. Wireless Commun.}, 
  title="{Deterministic Pilot Design and Channel Estimation for Downlink Massive MIMO-OTFS Systems in Presence of the Fractional Doppler}", 
  year={2021},
  volume={20},
  number={11},
  pages={7151-7165},
  doi={10.1109/TWC.2021.3081164}}

@BOOK{delay_Doppler_book,
  TITLE = "{Delay-Doppler Communications}",
  SUBTITLE = "{Principles and Applications}",
  AUTHOR = {Hong, Yi and Thaj, Tharaj and Viterbo, Emanuele},
  YEAR = {2022},
  PUBLISHER = {Elsevier},
}

@ARTICLE{tharaj_rake_MRC_journal,
  author={Thaj, Tharaj and Viterbo, Emanuele},
  journal={IEEE Trans. Veh. Tech.}, 
  title="{Low Complexity Iterative Rake Decision Feedback Equalizer for Zero-Padded OTFS Systems}", 
  year={2020},
  volume={69},
  number={12},
  pages={15606-15622},
  doi={10.1109/TVT.2020.3044276}}

@ARTICLE{tharaj_OTSM,
  author={Thaj, Tharaj and Viterbo, Emanuele and Hong, Yi},
  journal={IEEE Transactions on Wireless Communications}, 
  title="{Orthogonal Time Sequency Multiplexing Modulation: Analysis and Low-Complexity Receiver Design}", 
  year={2021},
  volume={20},
  number={12},
  pages={ 7842-7855},
  doi={10.1109/TWC.2021.3088479}}

@ARTICLE{tharaj_universal_MRC,
  author={Thaj, Tharaj and Viterbo, Emanuele and Hong, Yi},
  journal={IEEE Access}, 
  title="{General I/O Relations and Low-Complexity Universal MRC Detection for All OTFS Variants}", 
  year={2022},
  volume={10},
  number={},
  pages={96026-96037},
  doi={10.1109/ACCESS.2022.3204999}}

@INPROCEEDINGS{Ravi_Radar,
  author={Raviteja, P. and  Phan, K.T. and Hong, Y. and Viterbo, E.},
  booktitle={2019 IEEE Radar Conference (RadarConf)}, 
  title="{Orthogonal Time Frequency Space (OTFS) Modulation Based Radar System}", 
  year={2019},
  address={Boston, MA, USA },
  volume={},
  number={},
  pages={1-6},
  doi={10.1109/RADAR.2019.8835764}}

@ARTICLE{Thomas_ChaEst,
  author={Thomas, A. and Deka, K. and Raviteja, P. and Sharma, S.},
  journal={ IEEE Transactions on Communications}, 
  title="{Convolutional Sparse Coding Based Channel Estimation for OTFS-SCMA in Uplink}", 
  year={2022},
  volume={70},
  number={8},
  pages={5241-5257},
  doi={10.1109/TCOMM.2022.3182402}}

@ARTICLE{embedded_pilot_raviteja,
  author={Raviteja, P. and Phan, Khoa T. and Hong, Yi},
  journal={IEEE Trans. Veh. Tech.}, 
  title="{Embedded Pilot-Aided Channel Estimation for OTFS in Delay–Doppler Channels}", 
  year={2019},
  volume={68},
  number={5},
  pages={4906-4917},
  doi={10.1109/TVT.2019.2906357}}

@ARTICLE{Mishra_superimposed_pilots_ch_est_2022,
  author={Mishra, Himanshu B. and Singh, Prem and Prasad, Abhishek K. and Budhiraja, Rohit},
  journal={IEEE Trans Wireless Commun.}, 
  title="{OTFS Channel Estimation and Data Detection Designs With Superimposed Pilots}", 
  year={2022},
  volume={21},
  number={4},
  pages={2258-2274},
  doi={10.1109/TWC.2021.3110659}}

@ARTICLE{Shen_OMP_2019,
  author={Shen, Wenqian and Dai, Linglong and An, Jianping and Fan, Pingzhi and Heath, Robert W.},
  journal={IEEE Trans. Sig. Proc.}, 
  title="{Channel Estimation for Orthogonal Time Frequency Space (OTFS) Massive MIMO}", 
  year={2019},
  volume={67},
  number={16},
  pages={4204-4217},
  doi={10.1109/TSP.2019.2919411}}

@ARTICLE{Zhao_SBL_2020,
  author={Zhao, Lei and Gao, Wen-Jing and Guo, Wenbin},
  journal={IEEE Commun. Lett.}, 
  title="{Sparse Bayesian Learning of Delay-Doppler Channel for OTFS System}", 
  year={2020},
  volume={24},
  number={12},
  pages={2766-2769},
  doi={10.1109/LCOMM.2020.3021120}}

@ARTICLE{Suraj_BSBL_2021,
  author={Srivastava, Suraj and Singh, Rahul Kumar and Jagannatham, Aditya K. and Hanzo, Lajos},
  journal={IEEE Trans. Veh. Tech.}, 
  title="{Bayesian Learning Aided Sparse Channel Estimation for Orthogonal Time Frequency Space Modulated Systems}", 
  year={2021},
  volume={70},
  number={8},
  pages={8343-8348},
  doi={10.1109/TVT.2021.3096432}}

@ARTICLE{Liu_uplink_MIMO_ch_est_2020,
  author={Liu, Yushan and Zhang, Shun and Gao, Feifei and Ma, Jianpeng and Wang, Xianbin},
  journal={IEEE Journal Sel. Areas Commun.}, 
  title="{Uplink-Aided High Mobility Downlink Channel Estimation Over Massive MIMO-OTFS System}", 
  year={2020},
  volume={38},
  number={9},
  pages={1994-2009},
  doi={10.1109/JSAC.2020.3000884}}

@INPROCEEDINGS{Chocks_multi_user_MIMO_ch_est_2020,
  author={Rasheed, O K. and Surabhi, G. D. and Chockalingam, A.},
  booktitle={Proc. IEEE 91st Veh. Tech. Conf.}, 
  title="{Sparse Delay-Doppler Channel Estimation in Rapidly Time-Varying Channels for Multiuser OTFS on the Uplink}", 
  year={2020},
  address={ Antwerp, Belgium},
  volume={},
  number={},
  pages={1-5},
  doi={10.1109/VTC2020-Spring48590.2020.9128497}}

@INPROCEEDINGS{jesbin_superimposed,
  author={Jesbin, Fathima and Rao Mattu, Sandesh and Chockalingam, A.},
  booktitle={2023 IEEE Wireless Communications and Networking Conference (WCNC)}, 
  title="{Sparse Superimposed Pilot Based Channel Estimation in OTFS Systems}", 
  year={2023},
  volume={},
  number={},
  pages={1-6},
  doi={10.1109/WCNC55385.2023.10118899}}

@INPROCEEDINGS{learning_choks,
  author={Mattu, Sandesh Rao and Chockalingam, A.},
  booktitle={2022 IEEE 96th Vehicular Technology Conference (VTC2022-Fall)}, 
  title="{Learning based Delay-Doppler Channel Estimation with Interleaved Pilots in OTFS}", 
  year={2022},
  volume={},
  number={},
  pages={1-6},
  doi={10.1109/VTC2022-Fall57202.2022.10012974}}

@ARTICLE{LiXiangjun2025SBLseg,
  author={Li, Xiangjun and Liang, Yu and Zhou, Zhengchun and Fan, Pingzhi},
  journal={IEEE Communications Letters}, 
  title="{Fractional Delay-Doppler Channel Estimation for OTFS Systems Based on Segmentation Technique and Sparse Bayesian Learning}", 
  year={2025},
  volume={29},
  number={5},
  pages={1067-1071},
  doi={10.1109/LCOMM.2025.3553831}}

@ARTICLE{SaifChok2025interleavedCE,
  author={Jayachandran, Jinu and Khan, Imran Ali and Mohammed, Saif Khan and Hadani, Ronny and Chockalingam, Ananthanarayanan and Calderbank, Robert},
  journal={IEEE Transactions on Vehicular Technology}, 
  title="{Zak-OTFS with Interleaved Pilots to Extend the Region of Predictable Operation}", 
  year={2025},
  volume={},
  number={},
  pages={1-15},
  doi={10.1109/TVT.2025.3579394}}

@ARTICLE{YouGe2025OTFSvariants,
  author={Deng, Qinwen and Ge, Yao and Ding, Zhi},
  journal={IEEE Communications Surveys $\&$ Tutorials}, 
  title="{A Unifying View of OTFS and Its Many Variants}", 
  year={2025},
  volume={},
  number={},
  pages={1-1},
  doi={10.1109/COMST.2025.3542467}}

@ARTICLE{PriyaLargeDelay2024,
  author={Priya, Preety and Hong, Yi and Viterbo, Emanuele},
  journal={IEEE Transactions on Wireless Communications}, 
  title="{OTFS Channel Estimation and Detection for Channels With Very Large Delay Spread}", 
  year={2024},
  volume={23},
  number={9},
  pages={11920-11930},
  doi={10.1109/TWC.2024.3386160}}

@ARTICLE{RIS_OTFS_Yao,
  author={Li, Muye and Zhang, Shun and Ge, Yao and Gao, Feifei and Fan, Pingzhi},
  journal={IEEE Transactions on Communications}, 
  title="{Joint Channel Estimation and Data Detection for Hybrid RIS Aided Millimeter Wave OTFS Systems}", 
  year={2022},
  volume={70},
  number={10},
  pages={6832-6848},
  doi={10.1109/TCOMM.2022.3199019}}

@ARTICLE{Doppler_squint,
  author={Duan, Mingming and Zhang, Pengfei and Zhang, Shun and Ge, Yao and Dobre, Octavia A. and Yuen, Chau},
  journal={IEEE Transactions on Communications}, 
  title="{Channel Estimation and Hybrid Precoding for Massive MIMO-OTFS System With Doubly Squint}", 
  year={2025},
  volume={73},
  number={10},
  pages={9388-9404},
  doi={10.1109/TCOMM.2025.3562357}}

    \begin{IEEEbiography}[{\includegraphics[width=1in,height=1.25in,clip,keepaspectratio]{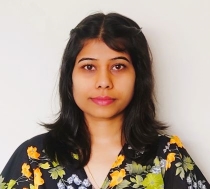}}]{Preety Priya} (GS'18--M'22) is currently an Assistant Professor at the Department of Electronics Engg., IIT (ISM) Dhanbad, India. She received her M.Tech. degree in telecommunication systems engineering in 2014, and Ph.D. degree in 2022 from the Indian Institute of Technology Kharagpur. From 2024 to 2025 she was an Assistant Professor at the Department of Electronics and Communication Eng., National Institute of Technology Calicut, India. Prior to that from 2022 to 2024 she was a Research Fellow with the Department of Electrical and Computer Systems Engineering, Monash University, Melbourne, Australia. From 2014 to 2016, she worked as an Assistant Professor with Kalinga Institute of Industrial Technology, Bhubaneswar, India.  She was awarded the Qualcomm Innovation Fellowship from Qualcomm in 2017. Her research interests include physical layer design of next generation wireless communications, OFDM, OTFS, ISAC, sparse signal processing, and nonlinear estimation. 
\end{IEEEbiography}
%%%%%%%%%%%%%

\begin{IEEEbiography}[{\includegraphics[width=1in,height=1.2in,clip]{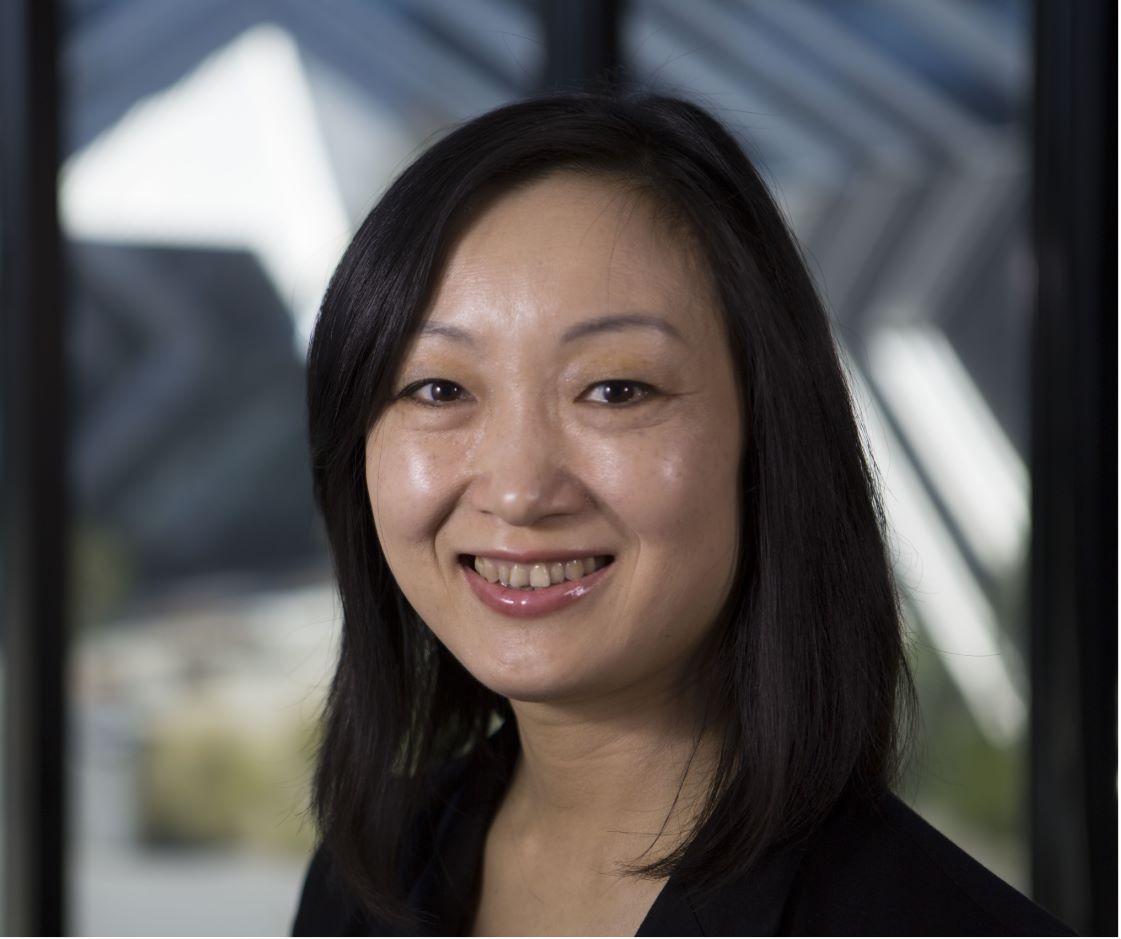}}]{Yi Hong}(S'00--M'05--SM'10)
is currently an associate professor at the Department of Electrical and Computer Systems Eng.,
Monash University, Melbourne, Australia.
She obtained her Ph.D. degree in Electrical Engineering and Telecommunications 
from the University of New South Wales (UNSW), Sydney, and received   
the {\em NICTA-ACoRN Earlier Career Researcher Award} at the {\em Australian Communication
Theory Workshop}, Adelaide, Australia, 2007. She served on the Australian Research Council College of Experts (2018-2020).
 
Prof Hong is currently an Associate Editor (AE) for the {\em IEEE Transactions on Green Communications and Networking}, 
and was the AE for the {\em IEEE Wireless Communication Letters} and {\em Transactions on Emerging Telecommunications Technologies (ETT)}.
She served as the Tutorial Chair of {\em IEEE International Symposium on Information Theory}, Melbourne, 2021,
and the General Co-Chair for the {\em OTFS workshops} in the {\em 2019-22 IEEE Communications Conferences},  
the {\em OTFS workshop} in the {\em IEEE Vehicular Technology Conference-Fall}, 2021, and the {\em IEEE Information Theory Workshop}, Hobart, 2014.
Her research interests include communication theory, coding and information theory with applications to telecommunication engineering.
\end{IEEEbiography}
%%%%%%%%%%%%%%
\begin{IEEEbiography}[{\includegraphics[width=1in,height=1.25in,clip,keepaspectratio]{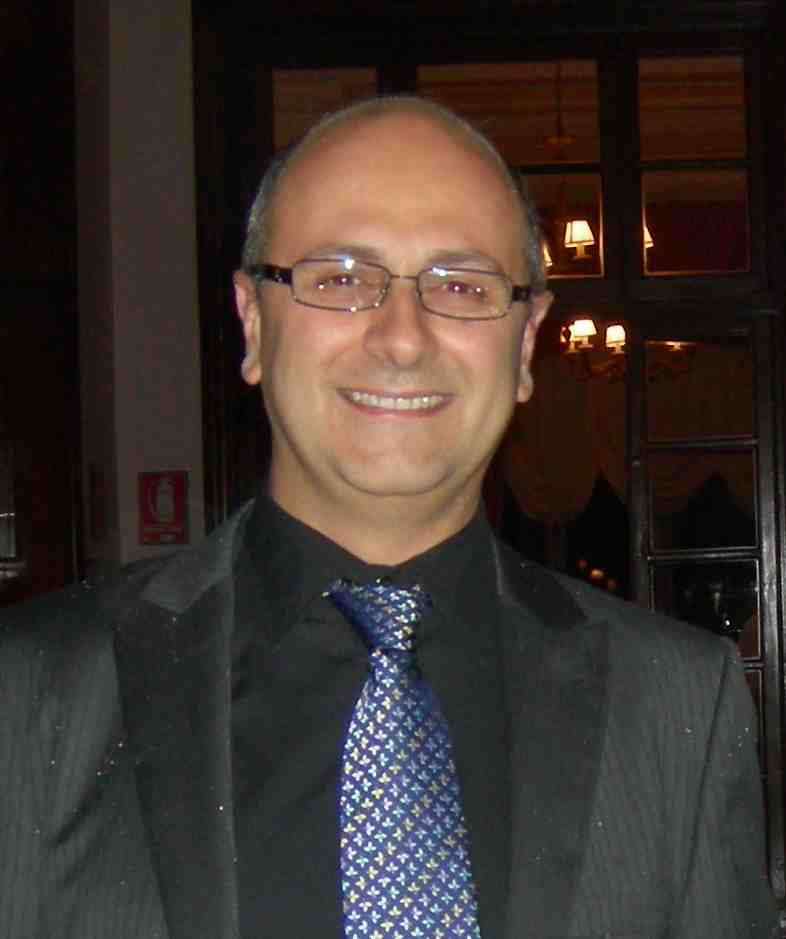}}]{Emanuele Viterbo}(M'95--SM'04--F'11) is currently a professor in the ECSE Department and an Associate Dean in Graduate Research at Monash University, Melbourne, Australia. He received his Ph.D. in 1995 in Electrical Engineering, from the Politecnico di Torino, Torino, Italy. From 1990 to 1992 he was with the European Patent Office, The Hague, The Netherlands, as a patent examiner in the field of dynamic recording and error-control coding. Between 1995 and 1997 he held a post-doctoral position in the Dipartimento di Elettronica of the Politecnico di Torino. In 1997-98 he was a post-doctoral research fellow in the Information Sciences Research Center of AT T Research, Florham Park, NJ, USA. From 1998-2005, he worked as Assistant Professor and then Associate Professor, in Dipartimento di Elettronica at Politecnico di Torino. From 2006-2009, he worked in DEIS at University of Calabria, Italy, as a Full Professor. Prof. Emanuele Viterbo is an ISI Highly Cited Researcher since 2009. He is an Associate Editor of  \emph{IEEE Trans. on Information Theory}, \emph{European Transactions on Telecommunications, and Journal of Communications and Networks}, and a Guest Editor of \emph{IEEE J. of Selected Topics in Signal Processing: Special Issue Managing Complexity in Multiuser MIMO Systems}.
Prof. Emanuele Viterbo was awarded a NATO Advanced Fellowship in 1997 from the Italian National Research Council. His main research interests are in lattice codes for the Gaussian and fading channels, algebraic coding theory, algebraic space-time coding, digital terrestrial television broadcasting, digital magnetic recording, and irregular sampling.
\end{IEEEbiography}
%%%%%%%%%%%%%%%%%%%%%%%%%%%%%%
%%%%%%%%%%%%%%
\end{document}